\newif\ifspie
\spiefalse

\ifspie
\documentclass[12pt]{spieman}
\def\basefigwidth{0.6\linewidth}
\else
\documentclass[letter,apj]{emulateapj-rtx4}
\def\basefigwidth{\linewidth}
\fi

\usepackage{graphicx}
\usepackage{enumerate}
\usepackage{amssymb, amsmath}
\usepackage{mathtools}
\usepackage{natbib}
\usepackage{color}
\usepackage{ulem}
\usepackage{soul}
\usepackage{bm}
\usepackage{rotating}
\usepackage{setspace}
\usepackage{tocloft}

\usepackage{tikz}
\usetikzlibrary{shapes,arrows}
\tikzstyle{block} = [draw, fill=white, rectangle, 
    minimum height=3em, minimum width=6em,align=center]
\tikzstyle{input} = [node distance=1cm]
\tikzstyle{output} = [node distance=1cm]
\tikzset{
  font={\fontsize{10pt}{12}\selectfont}}

\ifspie
\newcommand\apjl{ApJ}%
\newcommand\apjs{ApJS}%
\newcommand\ao{Appl.~Opt.}%
\newcommand\aap{A\&A}%
\else
\fi

\ifspie

\cftpagenumbersoff{figure}
\cftpagenumbersoff{table} 
\else
\fi

\begin{document}

\title{Data Reduction Pipeline for the CHARIS Integral-Field Spectrograph I: Detector Readout Calibration and Data Cube Extraction}

\ifspie
\else

\newcommand{\ucsb}{1}
\newcommand{\nasa}{2}
\newcommand{\stanford}{3}
\newcommand{\princetonastro}{4}
\newcommand{\princetonmae}{5}
\newcommand{\caltech}{6}
\newcommand{\subaru}{7}
\newcommand{\rakuten}{8}
\newcommand{\todai}{9}
\newcommand{\naoj}{10}
\newcommand{\abc}{11}

\author{Timothy D.~Brandt\altaffilmark{\ucsb},
Maxime Rizzo\altaffilmark{\nasa},
Tyler Groff\altaffilmark{\nasa},
Jeffrey Chilcote\altaffilmark{\stanford},
Johnny P.~Greco\altaffilmark{\princetonastro},
N.~Jeremy Kasdin\altaffilmark{\princetonmae},
Mary Anne Limbach\altaffilmark{\princetonmae},
Michael Galvin\altaffilmark{\princetonmae},
Craig Loomis\altaffilmark{\princetonastro},
Gillian Knapp\altaffilmark{\princetonastro},
Michael W.~McElwain\altaffilmark{\nasa},
Nemanja Jovanovic\altaffilmark{\caltech},
Thayne Currie\altaffilmark{\subaru},
Kyle Mede\altaffilmark{\rakuten},
Motohide Tamura\altaffilmark{\todai, \naoj, \abc},
Naruhisa Takato\altaffilmark{\subaru}, and
Masahiko Hayashi\altaffilmark{\naoj}
}

\altaffiltext{\ucsb}{Department of Physics, University of California, Santa Barbara, Santa Barbara, CA, USA}
\altaffiltext{\nasa}{NASA Goddard Space Flight Center, Greenbelt, MD, USA}
\altaffiltext{\stanford}{Department of Physics, Stanford University, Palo Alto, CA, USA}
\altaffiltext{\princetonastro}{Department of Astrophysical Sciences, Princeton University, Princeton, NJ, USA}
\altaffiltext{\princetonmae}{Department of Mechanical and Aerospace Engineering, Princeton University, Princeton, NJ, USA}
\altaffiltext{\caltech}{California Institute of Technology, Pasadena, CA, USA}
\altaffiltext{\subaru}{Subaru Telescope, National Astronomical Observatory of Japan, Hilo, HI, USA}
\altaffiltext{\rakuten}{Rakuten, Inc., Setagaya, Tokyo, Japan}
\altaffiltext{\todai}{Department of Astronomy, University of Tokyo, Hongo, Tokyo, Japan}
\altaffiltext{\naoj}{National Astronomical Observatory of Japan, Mitaka, Tokyo, Japan}
\altaffiltext{\abc}{Astrobiology Center of NINS, Osawa, Mitaka, Tokyo, Japan}

\fi

\begin{abstract}
We present the data reduction pipeline for CHARIS, a high-contrast integral-field spectrograph for the Subaru Telescope.  The pipeline constructs a ramp from the raw reads using the measured nonlinear pixel response, and reconstructs the data cube using one of three extraction algorithms: aperture photometry, optimal extraction, or $\chi^2$ fitting.  We measure and apply both a detector flatfield and a lenslet flatfield and reconstruct the wavelength- and position-dependent lenslet point-spread function (PSF) from images taken with a tunable laser.  We use these measured PSFs to implement a $\chi^2$-based extraction of the data cube, with typical residuals of $\sim$5\% due to imperfect models of the undersampled lenslet PSFs.  The full two-dimensional residual of the $\chi^2$ extraction allows us to model and remove correlated read noise, dramatically improving CHARIS' performance.  The $\chi^2$ extraction produces a data cube that has been deconvolved with the line-spread function, and never performs any interpolations of either the data or the individual lenslet spectra.  The extracted data cube also includes uncertainties for each spatial and spectral measurement.  CHARIS' software is parallelized, written in Python and Cython, and freely available on github with a separate documentation page.  Astrometric and spectrophotometric calibrations of the data cubes and PSF subtraction will be treated in a forthcoming paper.
\end{abstract}


\ifspie
{\noindent \footnotesize\textbf{*}Timothy D.~Brandt,  \linkable{tbrandt@ias.edu} }
\begin{spacing}{2} 
\else
\fi

\section{Introduction} \label{sec:intro}

Due to the advent of large format detectors, integral-field spectrographs (IFSs) have become an increasingly popular class of astronomical instrumentation.  IFSs are hybrids of traditional imaging cameras and slit spectrographs: they obtain a spectrum from each spatial element in a two-dimensional field-of-view for an $(x, y, \lambda)$ data cube.  The first realization of an IFS used a bundle of fibers to create a pseudo-slit \citep{Vanderriest_1980}, while TIGER \citep{Bacon+Baranne+Courtes+etal_1995} was the first IFS to use a lenslet array.  Modern IFSs generally use either fiber bundles \citep[MaNGA,][]{Drory+MacDonald+Bershady+etal_2015} or image slicers \citep[SINFONI, MUSE, NIRSpec][]{Eisenhauer+Abuter+Bickert+etal_2003, Bacon+Accardo+Adjali+etal_2010,Bagnasco+Kolm+Ferruit+etal_2007} 
to rearrange the field-of-view into a long pseudo-slit, or lenslet arrays \citep[OSIRIS,][]{Larkin+Barczys+Krabbe+etal_2006} to reimage spatial elements into small spots suitable for dispersion.

IFSs have become especially popular tools for high-contrast imaging.  Diffraction speckles have a different chromatic behavior from astrophysical sources; an IFS data cube can exploit this to achieve higher contrasts \citep{Sparks+Ford_2002,Marois+Correia+Galicher+etal_2014}.  An IFS also naturally enables the extraction of a planet or brown dwarf's spectrum, providing a probe of the object's temperature, chemistry and gravity \citep{McElwain+Metchev+Larkin+etal_2007, Barman+Macintosh+Konopacky+Marois_2011a, Konopacky+Barman+Macintosh+etal_2013, Hinkley+Pueyo+Faherty+etal_2013,Currie+Daemgen+Debes+etal_2014}.  IFSs combined with second-generation adaptive optics systems are now operational on Gemini South
\citep[GPI,][]{Macintosh+Graham+Palmer+etal_2008}, the VLT \citep[SPHERE,][]{Claudi+Turatto+Gratton+etal_2008}, and Palomar \citep[Project 1640,][]{Hinkley+Oppenheimer+Zimmerman+etal_2011}.  These new high-contrast instruments have recently discovered and characterized the low-mass companion to 51~Eri \citep{Macintosh+Graham+Barman+etal_2015}.
Future NASA mission studies such as Exo-C \citep{Stapelfeldt+Brenner+Warfield+etal_2014} and the WFIRST Coronagraph Instrument have baselined high contrast IFSs as their science cameras \citep{Spergel+Gehrels+Baltay+etal_2015,McElwain+Mandell+Gong+etal_2016}.

Data reduction and processing for IFSs has long presented problems.  The reduction pipeline for GPI \citep{Perrin+Maire+Ingraham+etal_2014,Perrin+Ingraham+Follette+etal_2016} is an ongoing, years-long effort partially built on legacy software from OSIRIS (whose pipeline also remains, to some degree, a work in progress).  This is the result of many complexities inherent in IFS data.  There are now two flatfields (one for the detector and one for the illumination of the fibers, lenslets, or sliced image plane).  For a lenslet-based IFS, the point-spread function (PSF) of the input optics and of the lenslets are both important.  The finite size of the lenslet PSFs means that neighboring spectra partially overlap one another; this must be corrected or accounted for during the extraction.  An IFS requires the wavelength and spectrophotometric calibrations of a spectrograph as well as the astrometric calibration of an imager.

This paper presents the data reduction pipeline for the CHARIS IFS on the Subaru telescope.  CHARIS, the Coronagraphic High Angular Resolution Imaging Spectrograph, is a lenslet-based near-infrared IFS located on the Nasmyth platform behind the adaptive optics systems AO188 \citep{Minowa+Hayano+Watanabe+etal_2010} and SCExAO \citep{Jovanovic+Martinache+Guyon+etal_2015}.  Section \ref{sec:charis_summary} summarizes the design and properties of the IFS, while the rest of the paper presents the software that extracts the data cube.  Section \ref{sec:readstoramp} discusses the construction of the pixel-by-pixel count rates from a sequence of raw reads, Section \ref{sec:calib} discusses our calibration procedure and associated data products, and Section \ref{sec:CubeExtraction} presents our algorithms for extracting the data cube.  Section \ref{sec:settings} summarizes the software's parameters and settings and Section \ref{sec:performance} shows the software's performance.  We discuss and conclude with Section \ref{sec:conclusions}.

This software package constructs the data cube and its inverse variance from a sequence of CHARIS reads.  We defer the necessary steps of image registration and spectrophotometric and astrometric calibration to a separate software package that is currently under development.  These steps depend on the SCExAO system in front of CHARIS and the calibrations changed when SCExAO was rebuilt in July of 2016; they also rely on a system of induced satellite spots that still must be manually controlled by the SCExAO team \citep{Jovanovic+Guyon+Martinache+etal_2015}.  These elements of the software, in addition to algorithms for angular and spectral differential imaging, will operate only on the data cubes produced by this pipeline and will be presented in a follow-up paper.

\section{The CHARIS Integral-Field Spectrograph} \label{sec:charis_summary}

CHARIS is a new high-contrast IFS for the Subaru Telescope.  Its scientific, conceptual, optical, and mechanical designs are summarized by \cite{McElwain+Brandt+Janson+etal_2012, Peters+Groff+Kasdin+etal_2012}, \cite{Galvin+Carr+Groff+etal_2014}, and \cite{Peters-Limbach+Groff+Kasdin+etal_2013}, respectively.  \cite{Groff+Chilcote+Kasdin+etal_2016} summarize laboratory testing performed after CHARIS was built but before it was transported to the summit.  This section briefly reviews the basic parameters and observing modes of CHARIS; we refer the reader to these other papers and to a forthcoming instrument paper for details.

CHARIS is a lenslet-based diffraction-limited spectrograph operating in the near-infrared.  Table \ref{tab:charis_params} summarizes its basic as-built properties.  CHARIS uses one of two prisms behind the lenslet array to disperse the light from each lenslet into a $\sim$30-pixel-long microspectrum.  The detector image consists of about $135 \times 135$ of these microspectra, each containing the light incident on a single $16.4$~mas square lenslet, for a $\sim$2$.\!\!''2 \times 2.\!\!''2$ field-of-view.  

\ifspie
\else
\begin{deluxetable}{lr}
\tablecaption{Basic CHARIS Parameters}
\tablehead{
    Parameter &
    Value
}
\startdata
Detector & $2048 \times 2048$ Hawaii2-RG \\
\# of Lenslets & $135 \times 135$ \\
Lenslet Size & 16.4 mas \\
Field-of-View & $2.\!\!''2 \times 2.\!\!''2$ \\
Wavelength Coverage & 1.15 - 2.38~$\mu$m \\
Microspectrum Length & $\sim$30 pixels \\
$R = \lambda/\delta \lambda$ (2 pixels) & $\sim$20 (low-res), $\sim$75 (high-res) \\
Available Modes & $J$, $H$, or $K$ at $R \sim 75$ \\
 & $J+H+K$ or ND\tablenotemark{$\dagger$} at $R \sim 18$ 
\enddata
\tablenotetext{$\dagger$}{Filter is ND3 ($10^{-3}$ transmission) from 1.15 to 2.4~$\mu$m, opaque at other wavelengths.}
\label{tab:charis_params}
\end{deluxetable}

\fi

CHARIS offers five observing modes, three with a high-resolution prism and two with a low resolution prism.  We measure its as-built spectral resolution $R$ using the definition
\begin{equation}
R = \lambda/\delta \lambda = \left( 2 \frac{{\rm d} \ln \lambda}{{\rm d}x} \right)^{-1},
\end{equation}
with $x$ in pixels, so that the dispersion is a wavelength shift per two pixels (slightly larger than the full width at half maximum, or FWHM, of the lenslet PSFs).  With this definition, $R$ with the high-resolution prism varies from $\sim$85 at the short end of $J$ to $\sim$65 in the middle of $H$ to $\sim$85 at the long end of $K$.  The low-resolution prism has an $R$ that varies from about 18 to 22 across the $J$, $H$, and $K$ bands.  The high-resolution mode uses either a $J$, $H$, or $K$ band filter, while the low-resolution prism may be used either with a broadband or a neutral density (ND) filter.  The broadband filter has nearly unit transmission from 1.15 to 2.38~$\mu$m and sharp cutoffs toward both shorter and longer wavelengths.  The ND filter is ND3 (10$^{-3}$ transmission) between 1.15 and 2.38~$\mu$m and opaque at other wavelengths; it is inside the CHARIS dewar to reduce the thermal $K$-band background.  The ND filter is intended to allow observers to obtain unsaturated images of bright stars.  CHARIS' location at Subaru's Nasmyth platform gives it better stability than GPI or Project 1640, while its extremely flat dispersion across the $J$, $H$, and $K$ bands is unique among high-contrast IFSs.  CHARIS' use of pinholes on the lenslet array and its relatively generous spacing of microspectra result in very low spectral cross-talk \citep{Groff+Chilcote+Kasdin+etal_2016} and enable us to model and remove correlated read noise (Section \ref{subsec:readnoise}).

CHARIS is controlled by a Linux-based software and is integrated into the software environment of the observatory.  It has only three moving parts to be controlled during observations: a shutter, the five-slot filter wheel ($J$, $H$, $K$, broadband, and ND), and the three-position prism slider (low-resolution, high-resolution, and empty).  The only other command is to reset and read out the detector.  The rest of this paper presents the software for reconstructing the individual lenslet microspectra, i.e.~the data cube, from these reads.

\section{From the Reads to a Ramp} \label{sec:readstoramp}

CHARIS' detector is a Hawaii2-RG (H2RG), a HgCdTe CMOS device in which each pixel has its own amplifier.  The pixel is read out by measuring the voltage across it relative to a reference voltage in the system.  In CHARIS' configuration, 32 readout channels each read pixels at a rate of 100~kHz.  Including some down time as the readout proceeds to the next row of pixels, it takes 1.47 seconds to read out the full $2048\times2048$ pixel array.  Resetting the detector is done pixel-by-pixel, and also takes 1.47 seconds.  H2RG detectors have some generic shortcomings, including persistence after exposure to a bright source \citep{Smith+Zavodny+Rahmer+etal_2008} and $1/f$ read noise strongly correlated between readout channels \citep{Moseley+Arendt+Fixsen+etal_2010}.  

CHARIS always saves every one of its reads; this sequence of raw reads is then saved to disk.  The first step of the data extraction is to fit for the count rate at each pixel from the raw reads.  In the limit of a linear pixel response and the dominance of read noise over photon noise, the best-fit count rate $C_i$ in pixel $i$ may be derived from a $\chi^2$ fit to the sequence of reads $j$:
\begin{equation}
    \chi_i^2 = \sum_{{\rm reads}~j=1}^{N} \frac{\left( j\delta t \cdot C_i + b_i - F_{ij}\right)^2}{\sigma^2},
\end{equation}
where $\sigma^2$ is the variance from the read noise, $b_i$ is the pixel's reset value, $j \delta t$ is the time from reset at which pixel $i$ was read out for the $j^{\rm th}$ time, and $F_{ij}$ is the number of counts in read $j$ of pixel $i$.  Minimizing $\chi^2$ with respect to $C_i$ and $b_i$, we find that the best-fit count rate $C_i$ is given by a linear combination of the reads $j$, and the readout is called up-the-ramp \citep[UTR,][]{Fowler+Gatley_1990}:
\begin{equation}
C_i \delta t = \frac{12}{N^3 - N} \sum_{j=1}^{N} \left( j - \frac{N + 1}{2} \right) F_{ij},
\label{eq:uptheramp}
\end{equation}
where $N$ is the total number of reads.  Up-the-ramp, or a variant using only some of the available reads, is now commonly used to read out infrared arrays both on the ground and in space \citep{Rauscher+Fox+Ferruit+etal_2007, Finger+Dorn+Eschbaumer+etal_2008,Perrin+Maire+Ingraham+etal_2014,Dressel_2017}.

In the limit of uncorrelated read-noise, up-the-ramp improves on the signal-to-noise ratio of correlated double sampling (CDS), the normalized difference of the first and last reads, by a factor
\begin{equation}
\frac{{\rm SNR}_{\rm UTR}}{{\rm SNR}_{\rm CDS}} = \sqrt{\frac{N(N + 1)}{6(N - 1)}}.
\label{eq:rampsnr}
\end{equation}
If there are only two reads, i.e.~$N=2$, up-the-ramp and CDS are equivalent, and the ratio in Equation \eqref{eq:rampsnr} is unity.  It is never less than one, and increases asymptotically as $\sqrt{N/6}$, or $\sqrt{t_{\rm exp}/(9\,{\rm s})}$ assuming 1.5~seconds/read.  When photon noise dominates, up-the-ramp is asymptotically noisier than correlated double sampling by about 23\% \citep{Fixsen+Offenberg+Hanisch+etal_2000}.  This may be corrected by dynamically choosing different weights for each read at each pixel \citep[e.g.][]{Robberto_2014}, but such an approach is not easily compatible with our nonlinear fit (Section \ref{subsec:nonlinearity}).  In the high signal-to-noise regime we are limited by the fidelity of our models of the microspectra rather than by photon noise.

We compute a variance on a given pixel's count rate from the variance in the count rates of the reference pixels in its channel.  We then add photon noise assuming our configured gain of 2~$e^-$/count to be correct.  We have verified this gain by using a frame-to-frame scatter in count rate as a crude measure of shot noise.  In addition, we allow the user to enforce an additional error equal to a fixed fraction (with a suggested value of a few percent) of the count rate seen by each pixel.  This additional error accounts for our imperfect model of the microspectra and produces reduced $\chi^2$ values close to unity when fitting these models to the ramps; we add it in quadrature to the other errors.  The variances are saved as inverse variances so that bad pixels may be masked by giving them an inverse variance of zero.  Cosmic ray hits are rare in short CHARIS exposures, and we do not implement cosmic ray rejection within our ramps \citep{Fixsen+Offenberg+Hanisch+etal_2000,Offenberg+Fixsen+Rauscher+etal_2001}.  Instead, we mask pixels in which a single read exceeds that pixel's mean count rate by a factor of five and its mean read noise by a factor of ten.  Less than 0.03\% of pixels are masked in this way in a typical two minute exposure.

The CHARIS software generally uses the up-the-ramp coefficients in Equation \eqref{eq:uptheramp}.  The following subsections discuss the removal of an artifact in the first read of a CHARIS ramp and our handling of nonlinearity and saturation.  We then briefly discuss the read noise properties of our ramps.  In Section \ref{subsec:readnoise}, we will take advantage of the relatively low dimensionality of much of the read noise, fitting it out when extracting data cubes. 

\subsection{Correcting Artifacts in the First Read}

The first read in a CHARIS ramp is contaminated by an exponential decay of the reference voltage.  Figure \ref{fig:expdecay} shows the difference between the first and second reads of a dark frame, while the top panel of Figure \ref{fig:expdecay_corrected} shows the lower-left corner of a 17-read ramp taken in CHARIS' low-resolution mode without correcting for this decay.  Given CHARIS' baffling and low level of dark current, the mean count rate in a dark frame is much lower than the read noise; Figure \ref{fig:expdecay} shows a noisy decay to zero.  For the time series in Figure \ref{fig:expdecay}, the two-dimensional readout of each channel has been mapped back to one dimension assuming a pixel rate of 100~kHz and an 8~$\mu$s downtime between reading rows of pixels on the detector.  This exponential decay has a time constant of $\sim$21~ms and becomes negligible by 100--200~ms ($\sim$10\% of the 1.47~s full-frame readout time).  Similar artifacts have previously been noted in H2RGs and in related detectors \citep{Bacon+McMurtry+Pipher+etal_2004,Rauscher+Fox+Ferruit+etal_2007}.

\begin{figure}
    \centering
    \includegraphics[width=\basefigwidth]{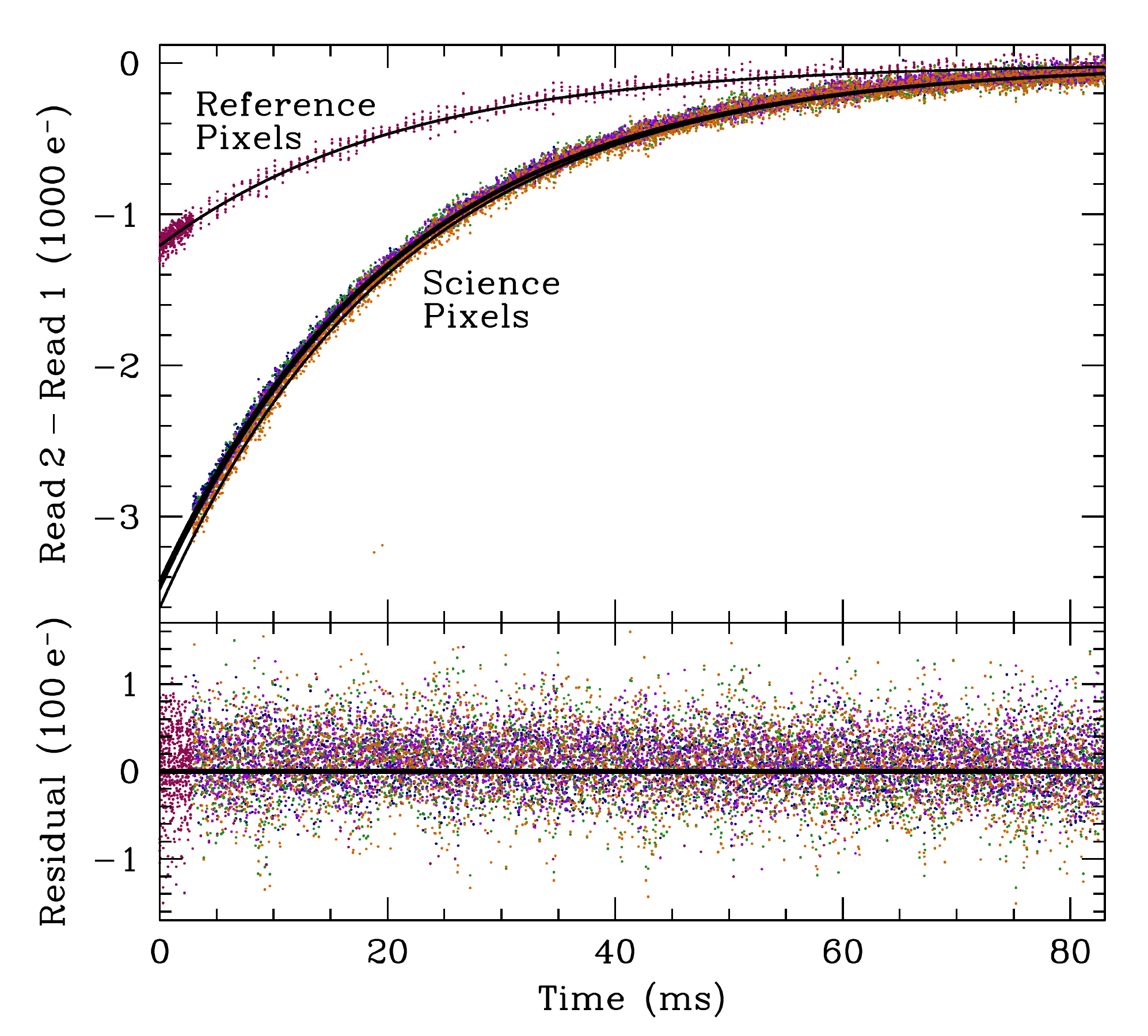}
    \caption{The exponential decay of the reference voltage in the first read, shown via the difference between the first and second reads in a dark frame.  Four channels are shown, in green, violet, blue, and orange, with the reference pixels in burgundy.  The solid black lines are exponential fits with a shared decay constant of about 21~ms; the lower panel shows the residuals.  The residuals deviate slightly from zero due to low-frequency noise on the detector.  Removing the exponential decay of the reference voltage allows us to use the first read in all CHARIS ramps.  }
    \label{fig:expdecay}
\end{figure}

The black lines in Figure \ref{fig:expdecay} take the form
\begin{equation}
y = A_i \exp \left[ -t/t_0 \right],
\end{equation}
where the time constant $t_0 = 21.1$~ms is the same for all of the lines and the $A_i$ are fitted separately to each readout channel $i$ (and to the reference pixels).  The lower panel shows the residuals, indicating a good fit with some remaining low-frequency read noise.  Removing the first-read artifact requires fitting 34 parameters in all: one time decay constant $t_0$, 32 amplitudes for the 32 readout channels, and an additional amplitude for the reference pixels.  We perform the fit as follows.

In a CHARIS ramp, the difference between the counts in the first and subsequent reads is the sum of the read noise, the exponential decay of the reference voltage, and the photon rates scaled by the gain.  We first estimate the photon count rates by fitting a ramp only to the second and subsequent reads, ignoring the contaminated first read.  We then use the fitted pixel-by-pixel count rates and reset values to obtain a modeled number of counts at the first read.  The difference between the actual and modeled counts in the first read is the sum of read noise and the reference voltage artifact that we wish to remove.  We must then perform a 34-parameter nonlinear fit.  Luckily, the fit is only nonlinear in a single parameter, the decay constant $t_0$.  Once the decay constant is fixed, the 34-parameter nonlinear optimization becomes 33 decoupled one-parameter linear optimizations.  By solving these linear problems for each value of $t_0$ we may reduce the problem to a one-dimensional nonlinear optimization for $t_0$.  We first guess the value of $t_0$ from the known behavior of the CHARIS detector, and then iteratively fit parabolas to converge to the best-fit decay constant.  This is equivalent to Newton-Raphson iteration on ${\rm d}\chi^2/{\rm d}t_0$.  Once we have the decay constant, the other 33 parameters may be obtained by straightforward linear optimizations.  The entire process for a typical CHARIS ramp takes a few hundred milliseconds on a laptop computer.

\begin{figure}
    \centering
    \includegraphics[width=\basefigwidth]{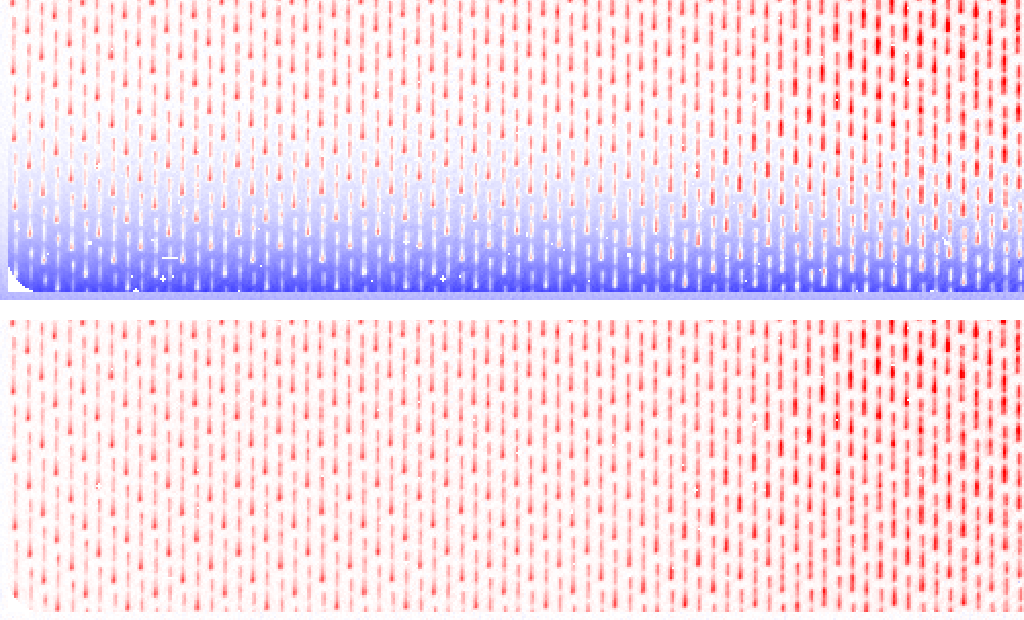}
    \caption{The lower-left corner of a 17-read ramp taken in CHARIS' broadband, low-resolution mode before (top) and after (bottom) removing the exponential decay of the reference voltage in the first read.  Fitting the reference voltage requires a negligible fraction of the information contained in the first read.  The read may then be used normally to increase the integration time on-source and reduce read noise.  The microspectra are visible as short vertical lines with gaps corresponding to the atmospheric absorption bands between the $J$ and $H$ and between the $H$ and $K$ bands.}
    \label{fig:expdecay_corrected}
\end{figure}

The lower panel of Figure \ref{fig:expdecay_corrected} shows the 17-read ramp of the top panel after removing the exponential decay of the reference voltage.  There are no longer any artifacts visible, and the first read may now be used to increase the integration time on-source and reduce the read noise.  The first read consists of $2048\times2048$ pixels, many orders of magnitude larger than the 34 parameters describing the reference voltage artifact: fitting out this artifact uses a negligible amount of the information available in the first read.

\subsection{Nonlinearity and Saturation} \label{subsec:nonlinearity}

As CHARIS' H2RG approaches saturation, its response becomes nonlinear.  The nonlinearity sets in gradually before the pixel's response drops sharply to zero.  The H2RG also suffers from bleeding into adjacent pixels.  After a pixel saturates, its four nearest neighbors see an immediate and compensating increase in their count rates, while its four next-nearest neighbors (the nearest along a diagonal) see a smaller, but still significant, increase.  Pixels that do not adjoin the saturated pixel show no significant increase in their count rates until an adjacent pixel saturates.  

\begin{figure}   
    \centering
    \includegraphics[width=\basefigwidth]{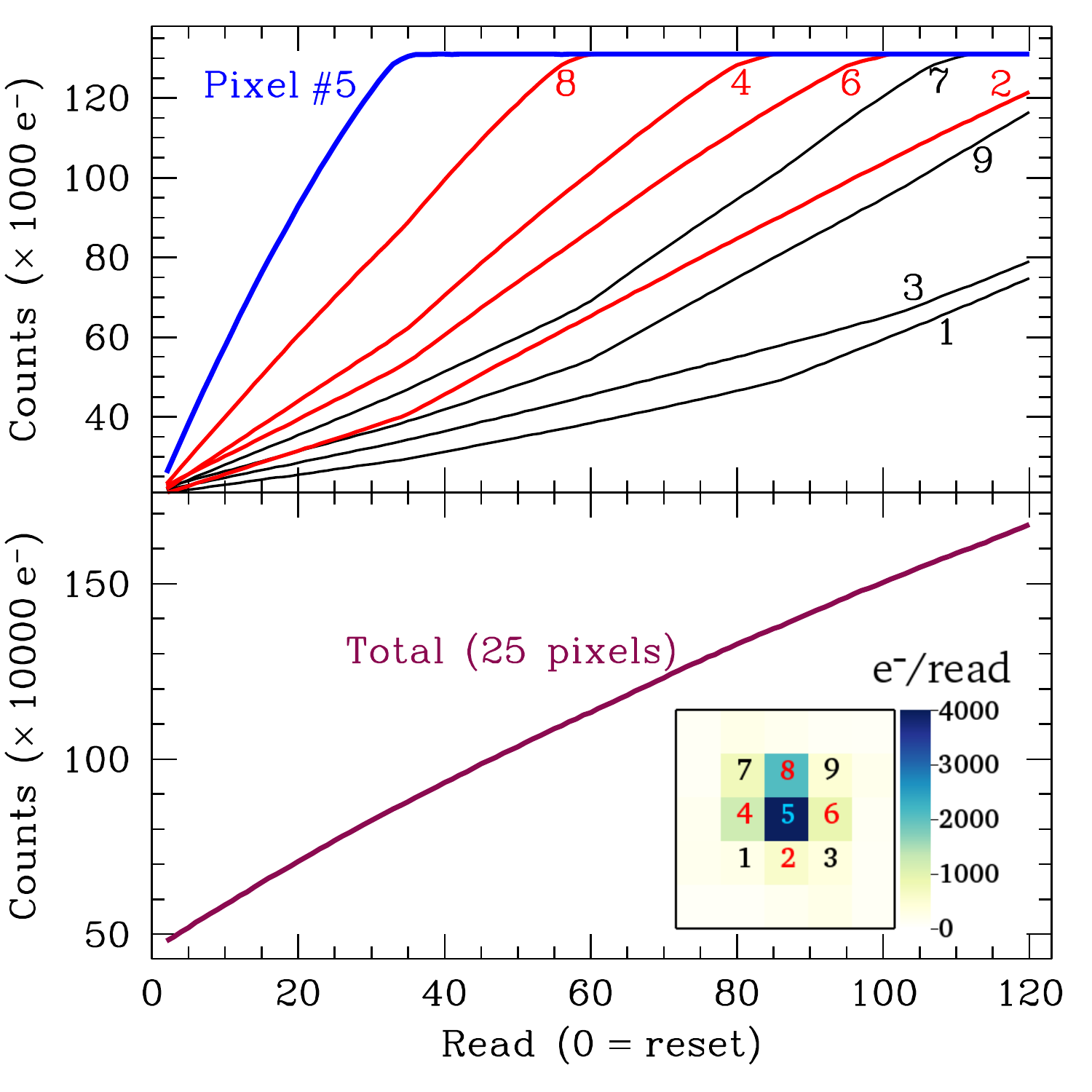}
    \caption{Leakage of electrons into neighboring pixels upon saturation.  Top panel: the actual counts (assuming a gain of 2) as the individual pixels shown in the inset saturate.  Neighboring pixels absorb the extra electrons upon saturation.  Lower panel: in spite of the saturation of pixels 4, 5, 6, 7, and 8 (accounting for nearly 70\% of the counts rate in the initial few reads), the total count rate in the $5 \times 5$ pixel box decreases by just $\sim$35\% between the first few and the last few reads.}
    \label{fig:bleeding}
\end{figure}

Figure \ref{fig:bleeding} shows the detector's response to saturation in a cutout of a 120-read ramp in which the most strongly illuminated pixels saturate in $\sim$30 reads.  The lines on the top panel are labeled with the pixel number; an inset in the lower panel provides the key.  The increase in count rates in the peripheral pixels closely approximates the lost counts from the saturated pixel(s).  The total instantaneous count rate in a $5 \times 5$ box decreases by just $\sim$35\% between the first few and the last few reads, despite the fact that saturated pixels account for $\sim$70\% of the photons in the initial reads.

The H2RG's saturation behavior means that the saturation of one pixel immediately corrupts the count rates of its eight nearest neighbors, but can be neglected for pixels that are farther away.  We fit for each pixel's count rate using only the reads for which neither the pixel in question, nor any of its immediate neighbors, has saturated.

We measure our H2RG's nonlinearity using a series of long exposures: a 1000-read ramp taken with the detector almost uniformly illuminated, and sixteen 120--160-read ramps with the detector sparsely illuminated by monochromatic light passing through the lenslets.  In both cases, we fit for each pixel's count rate using only the first $\sim$5\% of the reads.  We then compare the actual counts at subsequent reads to the expected count rates from the initial reads assuming perfect linearity.

\begin{figure*}[t]
    \centering
    \includegraphics[width=0.9\linewidth]{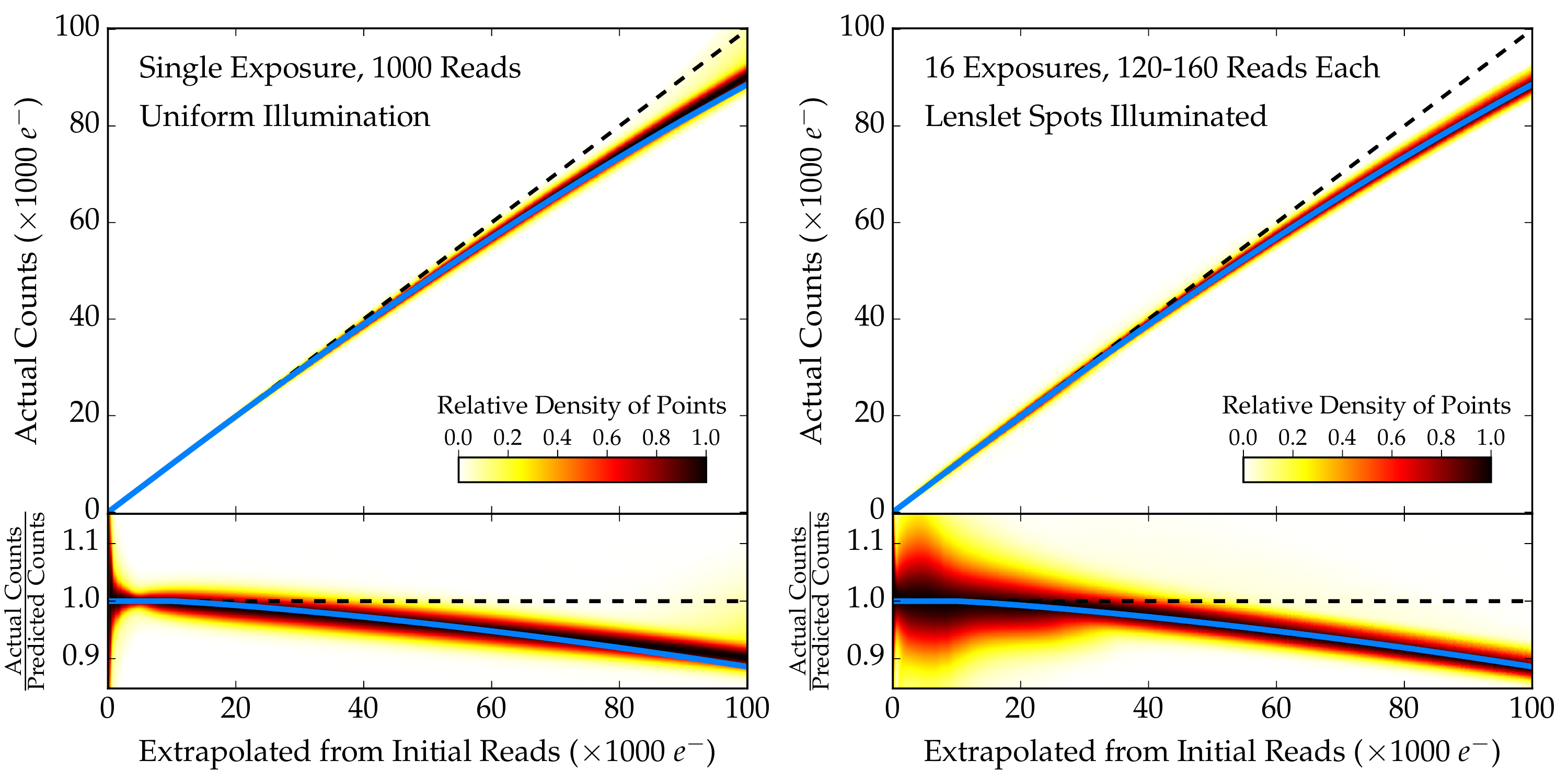}
    \caption{The nonlinear response of CHARIS' detector as measured in a 1000 read ramp taken before the full baffling was installed (left panel) and in a series of 120--160 read ramps taken with the detector sparsely illuminated by monochromatic light passing through the lenslet array.  The color shows the density of measurements normalized to the maximum density at a given extrapolated number of counts (x-coordinate); each plot contains data from several $10^9$ points.  
    The black dashed lines show a perfectly linear response, while the blue lines show our fiducial fit to the nonlinear response.  The frames differ slightly in the nonlinear response, at the level of  $\sim$1--2\%, as the pixels approach saturation ($\sim$100,000~$e^-$ above reset).  Our fit assumes the pixels to have a linear response up to 10,000~$e^-$ ($\sim$10\% of well capacity).  }
    \label{fig:nonlinearity}
\end{figure*}

Figure \ref{fig:nonlinearity} shows the density of points in actual vs.~predicted counts on the detector.  The density is independently normalized at each count rate ($x$-coordinate) to limit dynamic range; there are several billion points in each figure.  A perfectly linear detector should be nearly symmetric about the dashed line $y=x$ (biases from read and photon noise are $<$1\% at these count rates).  Our H2RG falls below the line, indicating a loss of sensitivity as the pixels approach well capacity.  The solid blue line shows our adopted nonlinear response: linear up to 10000~$e^-$, continuously matched to a cubic fit up to saturation.  Our method is distinct from \cite{Finger+Dorn+Eschbaumer+etal_2008}, who only use unsaturated reads to fit a ramp but assume the pixel response to remain linear.

Different ramps suggest nonlinear responses that differ by $\sim$1--2\% near saturation.  We make no attempt to explain or account for this in the data, but we do allow the user to add a fixed fraction of the count rate as an uncertainty.  This error term also accounts for imperfect modeling of the microspectra; an appropriate value (typically $\sim$5\%) gives a $\chi^2$ per pixel of order unity after fitting all of the two-dimensional microspectra.  This 5\% uncertainty is significantly larger than the $\sim$1\% uncertainty in the nonlinear response.

Our adopted pixel response is linear up to 10000~$e^-$, or $\sim$10\% of well capacity.  For pixels that remain below this value we therefore use the up-the-ramp fit described in the first paragraphs of Section \ref{sec:readstoramp}.  We individually fit pixels that exceed this threshold.  With a fixed nonlinear pixel response, the nonlinear fit has two free parameters: the initial count rate and the reset value.  The best-fit count rate $x$ minimizes
\begin{equation}
\chi^2[x,b] = \sum_j \frac{\left( F_j - f[x\cdot j] + b\right)^2}{\sigma^2},
\label{eq:nonlinear}
\end{equation}
where the sum is over the reads $j$, $b$ is the reset value, and we neglect photon noise.  For a linear detector, $f[x \cdot j] = x\cdot j$ with $x$ in units of counts/read and the best-fit $x$ is given by the usual up-the-ramp weights (Equation \eqref{eq:uptheramp}).  We exclude from the fit all pixels with a predicted count rate more than $>$10$^5$~$e^-$ above reset or with $>$60,000 raw counts ($1.2\times10^5$~$e^-$ at a gain of 2; values above 65535 raw counts cannot be represented by unsigned 16-bit integers).

In our case, $f[x \cdot j]$ is a nonlinear function of $x$, and the count rate may no longer be obtained as the weighted sum of the reads with the coefficients from Equation \eqref{eq:uptheramp}.  However, at fixed count rate, finding the best-fit reset value $b$ remains a linear problem, and the best $\chi^2$ at this count rate is trivial to compute.  We use this fact to reduce the two-dimensional nonlinear minimization in $x$ and $b$ to a one-dimensional nonlinear minimization in $x$ (the coefficients of the cubic function define the nonlinear response and remain fixed).  We begin with a guess from the up-the-ramp count rate and locally fit a parabola to $\chi^2[x]$; the vertex of the parabola is our next guess for the count rate.  As with our procedure to fit for the initial exponential decay of the reference voltage, this is equivalent to Newton-Raphson iteration on ${\rm d}\chi^2/{\rm d}x$.  The algorithm converges in only a few steps and requires a negligible amount of computation for a typical ramp.

\subsection{Read Noise Properties} \label{subsec:readnoiseprop}

CHARIS' H2RG detector has $1/f$ read noise that is correlated among the readout channels; this behavior is typical of H2RGs \citep{Moseley+Arendt+Fixsen+etal_2010}.  It may be suppressed by measuring by fitting out power on various timescales using interspersed reference pixels \citep{Moseley+Arendt+Fixsen+etal_2010}, using (nearly) unilluminated light-sensitive pixels \citep{Brandt+McElwain+Turner+etal_2013}, or using different weightings of the reference pixels at the detector edges \citep{Kubik+Barbier+Castera+etal_2014}.  In CHARIS, there is an additional component of read noise that is largely shared by alternating readout channels.  We read out our detector using 32 readout channels, each $2048 \times 64$ pixels in size.  The even and odd channels each have their own shared component of read noise: removing a single scaled template from all channels achieves only $\sim$half the noise suppression of using different templates for the even and odd channels.

The power spectrum of CHARIS' read noise spikes at a range of frequencies in the 1--10 kHz~range, most of which do not match any known frequencies of the system.  The noise is also highly variable with time, both in amplitude and in its power spectrum.  In lab tests during CHARIS's assembly, the shared read noise was comparable in power to the independent read noise in each channel.  On Subaru's Nasmyth platform, the correlated component of the read noise has gotten much worse: its variance can be more than ten times that of the independent read noise depending on the date and the readout channel.  We are currently investigating this source of noise and attempting to fix it in hardware.

\begin{figure*}
    \includegraphics[width=\linewidth]{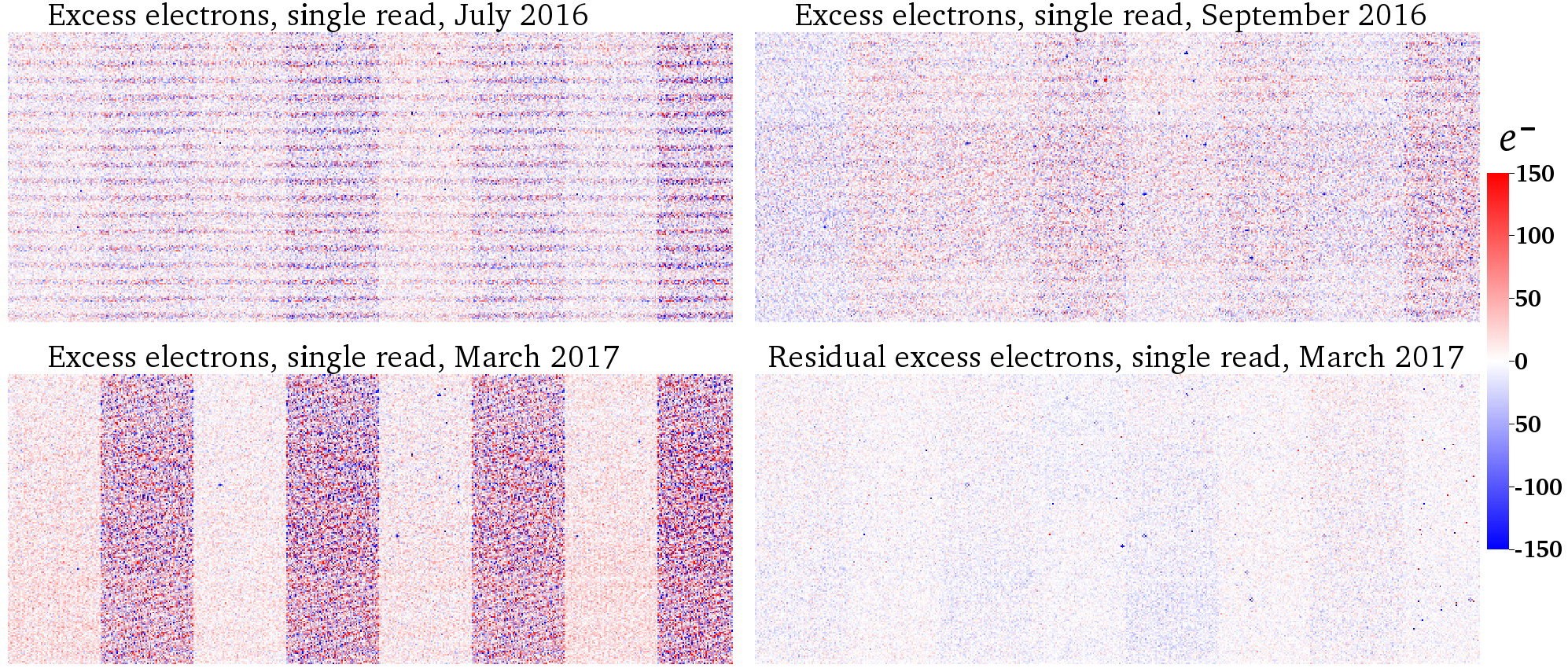}
    \caption{Sample realizations of the read noise on three different dates.  The read noise is computed as the difference between the actual number of counts at a single read and the expected number of counts based on the other reads in the ramp.  The July 2016, September 2016, and March 2017 images have root-mean-square noise of about 33\,$e^-$, 31\,$e^-$, and 53\,$e^-$, respectively.  The strong banding in July 2016 (much fainter but still visible in September) is modulation at a frequency of 120~Hz.  In March of 2017, the read noise shows a dramatic difference in amplitude between the even and odd readout channels.  Much of the noise in all of these images is in two patterns, one common to all of the even channels and one to all of the odd channels.  The lower-right panel shows the same March 2017 read as the lower-left panel, but with a scaled pattern subtracted from the even readout channels, and a different scaled pattern subtracted from the odd channels.  The read noise falls by a channel-dependent factor of 1.3 to 5.8 (the variance falls by a factor of $\sim$2 to 30), when this correlated read noise is removed; the overall root-mean-square falls from about 53\,$e^-$ to about 14\,$e^-$.  We discuss the removal of correlated read noise from CHARIS ramps in Section \ref{subsec:readnoise}.}
    \label{fig:readnoisesamples}
\end{figure*}

Figure \ref{fig:readnoisesamples} shows three representative realizations of CHARIS read noise on three widely separated dates.  In all cases, we have used a long ramp to predict the count rate at the second read of the ramp, and then used the difference between the actual and expected counts in the second read as a measure of the read noise.  The noise in a CDS image (a difference of two reads) would be $\sqrt{2}$ times the noise shown.  

Figure \ref{fig:readnoisesamples} shows that the read noise on Subaru's Nasmyth platform is a serious problem, with CDS-equivalent read noise $\gtrsim$100\,$e^-$ in some channels at some times, and that the read noise properties vary with time as CHARIS and neighboring instruments are moved and reconfigured.  The lower-right panel of Figure \ref{fig:readnoisesamples} demonstrates how much of this read noise is shared between alternating readout channels.  In this panel, we have constructed two read noise templates: one for the even channels, and a separate one for the odd channels.  We then rescaled the appropriate template to each channel and removed it.  This procedure reduced the read noise by a channel-dependent factor of 1.3 to 5.8.  If there were no correlated read noise, this procedure would only be expected to reduce the variance by $\sim$1/16, or the noise by $\sim$3\% (since sixteen channels were used to construct each template). 

Real CHARIS ramps are packed with microspectra, preventing us from simply removing the correlated read noise as described above.  However, the combination of our $\chi^2$ method of fitting the microspectra (Section \ref{subsec:chisqextract}) and the generous spacing of the spectra on the detector do enable us to achieve effective read noise suppression.  We discuss this in more detail in Section \ref{subsec:readnoise}.

\section{Calibrations for Cube Extraction using Monochromatic Flatfields}
\label{sec:calib}

A CHARIS image consists of microspectra arrayed in a grid on the detector (see Figure \ref{fig:expdecay_corrected}); these must be extracted into a data cube.  Each microspectrum is the integral of the monochromatic lenslet spots over the spectrum of light that that lenslet sees, convolved with the pixel response function.  Extracting the source spectrum requires knowing the pixel locations where each wavelength of light falls for each lenslet; this is the wavelength solution.  Optimal extraction and $\chi^2$ extraction, the two main algorithms included in the CHARIS data reduction software, also require knowledge of the monochromatic lenslet spots (the lenslet point-spread function, or lenslet PSF).  In the remainder of this paper we will use the terms lenslet PSF and PSFlet interchangeably.

In order to both derive the wavelength solution and measure the wavelength-dependent PSFlets, we inject a supercontinuum source with a narrowband tunable filter into an integrating sphere to uniformly illuminate CHARIS' lenslet array with monochromatic light.  The tunable filter has a width of 5~nm, for a spectral resolution of $\sim$300 at 1.6~$\mu$m, well in excess of CHARIS' resolving power in its high spectral resolution mode.  Such a calibration strategy would need to be revised for an instrument with higher spectral resolution, though a lamp with well-spaced emission lines might serve as an effective substitute for our tunable filter.  We have also integrated three narrowband filters, one each in the $J$, $H$, and $K$ bands, into SCExAO's optics.  With one of these filters in place, we may uniformly illuminate CHARIS's lenslet array with any infrared-bright lamp or even the twilight sky.

We break the CHARIS calibration procedure into two steps.  We first use our supercontinuum source and tunable filter to gradually step through wavelength.  This enables us to measure the locations of the lenslet PSFs on the detector and to derive the wavelength solution.  We also use these images to extract the position-dependent lenslet PSFs.  This first step, a full sequence of calibration images, is rarely (if ever) repeated.  As of publication, the pipeline includes calibration products derived from a July 2016 calibration sequence.  A new calibration sequence offers negligible improvements even one year and several cooling cycles later.  The second step is to use a narrowband flat to derive a correction to the wavelength solution once per night.  This section describes the process in detail, and summarizes the final calibration files that it produces.  

The typical user of CHARIS will not need to derive the full wavelength solution or the  monochromatic lenslet PSFs; we have calculated these and distribute them with the source code.  The user will, however, need to build the appropriate calibration files for a given observing mode from a single narrowband flat: thermal cycles of the instrument induce small shifts of the microspectra on the detector.  We include the script \verb|buildcal| for this purpose.  It takes as input the raw reads from a single narrowband flat and writes all of the necessary calibration files to the directory from which it was run.

\subsection{Full Calibration Sequences}

A full calibration sequence consists of connecting our supercontinuum source to a tunable filter, coupling it to an integrating sphere, and injecting the light into CHARIS to illuminate the lenslet array.  We then step through wavelength with a step size comparable to or slightly larger than CHARIS's spectral resolution.  This yields a series of about 10--15 images per observing mode with well-separated lenslet PSFs.  Figure \ref{fig:calsequence} shows these spectrally unresolved flatfields from 1.35~$\mu$m to 2.15~$\mu$m for a small subregion of the detector.

\begin{figure*}
    \centering
    \includegraphics[width=0.9\textwidth]{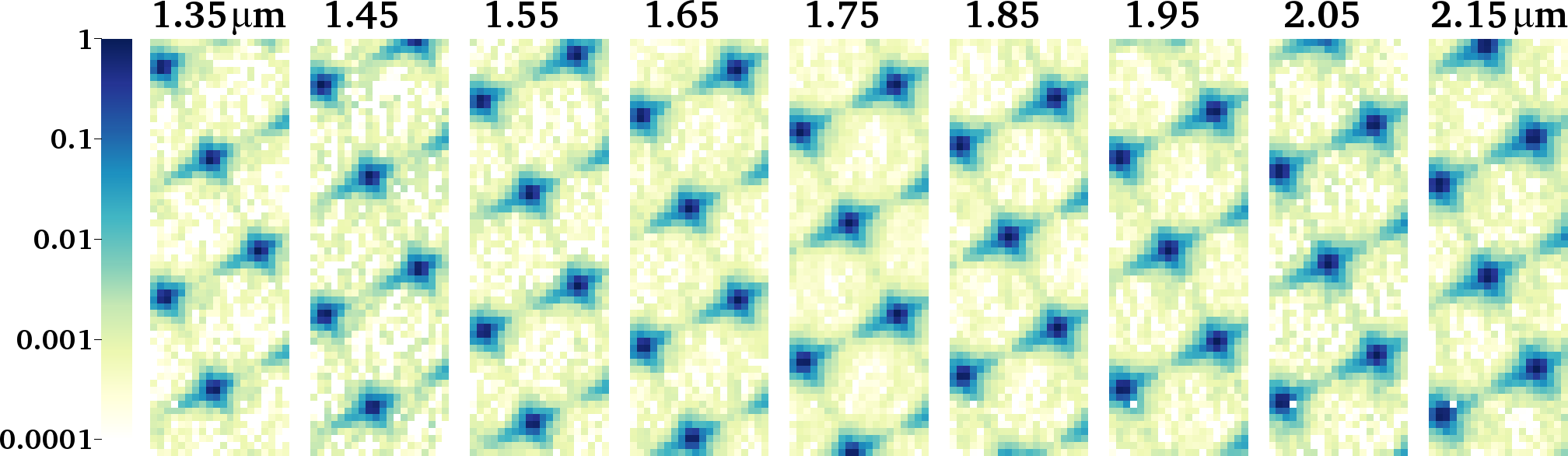}
    \caption{A partial calibration sequence through CHARIS's broadband mode showing the motion of the lenslet PSFs, or PSFlets, in the dispersion direction (toward the bottom of the page).  Only a small region of the detector is shown ($20\times 60$ pixels; a microspectrum is just over 30 pixels long), and the intensity is plotted on a logarithmic scale normalized to the peak of the brightest PSFlet in the region shown.  The diagonal spikes are due to diffraction from the square lenslets.}
    \label{fig:calsequence}
\end{figure*}

We use these calibration sequences to measure the \mbox{PSFlet} locations for each lenslet as a function of wavelength.  We assume a cubic polynomial mapping between integer lenslet coordinates $(i, j)$ and floating point pixel coordinates $(x_{ij}[\lambda], y_{ij}[\lambda])$, with the polynomial coefficients being functions of wavelength:
\begin{equation}
    \begin{Bmatrix}x_{ij}[\lambda]\\ y_{ij}[\lambda]\end{Bmatrix} = \sum_{m=0}^3\sum_{n=0}^{3-m} \begin{Bmatrix}a_{nm}[\lambda]\\ b_{nm}[\lambda]\end{Bmatrix} i^m j^n.
\end{equation}

We first lightly smooth our monochromatic images with a narrow Gaussian; we use the known lenslet pitch and rotation as our initial guesses for the linear coefficients of the polynomial.  We then maximize the sum of the interpolated intensities at the lenslet spot locations by adjusting the coefficients.  Once we have derived these coefficients for one wavelength, we estimate the offset in the dispersion direction for the next wavelength using a grid search.  We combine this new offset with the polynomial coefficients from the previous step to form the initial guess for this new optimization. We proceed to derive the cubic polynomial transformation from lenslet to detector coordinates for all wavelength steps.

For any wavelength of interest, we can now compute all of the coefficients of the lenslet-detector transformation polynomial (and hence the full wavelength solution) by fitting a cubic polynomial as a function of $\log \lambda$ to each coefficient.  We have derived these full wavelength solutions for each CHARIS observing mode using calibration sequences taken in July 2016 and distribute them as part of the CHARIS software package.  

\begin{figure}
    \centering
    \includegraphics[width=\basefigwidth]{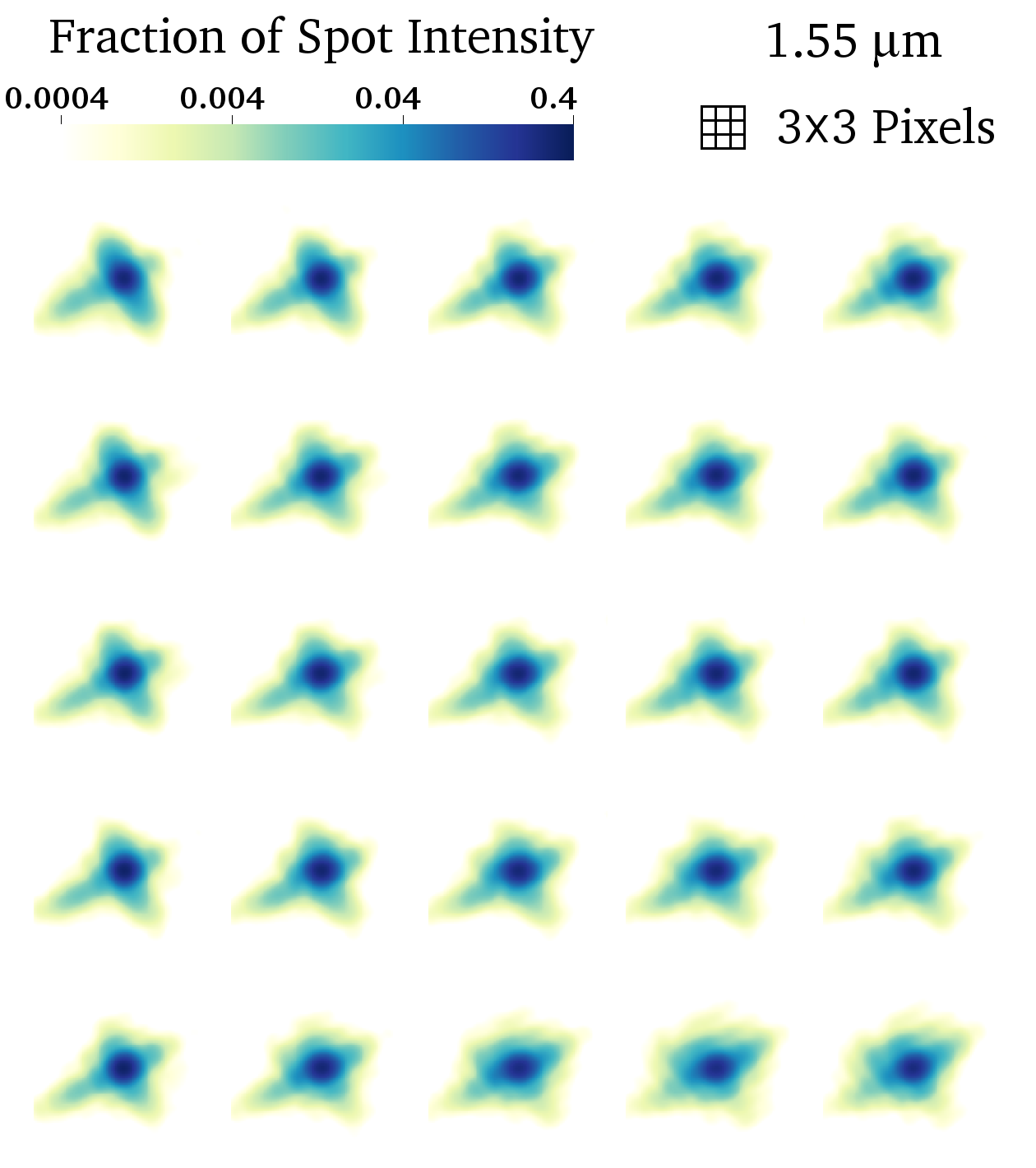}
    \caption{Oversampled 1.55~$\mu$m PSFlets reconstructed over 25 subregions of the detector.  The observed PSFlets are resampled by the integer pixel grid of the detector; a $3\times 3$ pixel grid in the upper-right shows the pixel size.  The PSFlets are normalized to unit intensity after resampling.  The oversampled PSFlets allow us to construct the pixellated microspectra corresponding to monochromatic and broadband light incident on the lenslet array.  The crosses are diffraction spikes from the square lenslets, while the shape variations show the focus changing across the field.}
    \label{fig:hires_psflets}
\end{figure}

Our next step is to reconstruct the lenslet PSFs.  For wavelengths $\lesssim$2~$\mu$m, these are undersampled by CHARIS's H2RG.  We take an approach very similar to \cite{Anderson+King_2000} for {\it Hubble} and to \cite{Ingram+Ruffi+Perrin+etal_2014} and \cite{Draper+Marois+Wolff+etal_2014} for GPI.  By deriving the wavelength solution, we already have the location of each lenslet PSF's centroid and can place it on an oversampled grid.  The lenslet PSFs are not spaced by an integer number of pixels nor by a ratio of small integers; as a result, they populate an oversampled PSFlet reasonably densely.  We iteratively construct a slightly smoothed PSFlet and fit for each PSFlet's normalization to refine the smoothed template.  Finally, we deconvolve with the smoothing kernel using the Richardson-Lucy algorithm \citep{Lucy_1974} to produce our final oversampled PSFlets.  We measure these oversampled PSFlets at about ten wavelengths in each observing mode.

Figure \ref{fig:hires_psflets} shows our resulting oversampled PSFlets in 25 subregions of the detector for 1.55~$\mu$m monochromatic light.  We use bilinear interpolation to estimate the PSFlet between the centers of regions, and assume the PSFlets to remain constant from the centers of the outer regions to the edges of the detector.  The convolution and deconvolution are not completely equivalent because some subpixel offsets are sampled more than others; this could lead to a systematic underestimation of the PSFlet width.  Random errors in the PSFlet centers would push in the other direction, leading to a systematic overestimation of PSFlet width.  An inspection of the residuals after our $\chi^2$ fitting of the microspectra (Section \ref{subsec:chisqextract}) indicates that these errors are negligible.  Our use of only 25 images to represent PSFlet variation over the detector (motivated by the need to average over a large number of lenslets) is probably a bigger source of error.  

As of publication, the pipeline includes one set of calibration files derived from calibration sequences taken in July 2016.  We have been unable to conclusively measure any differences in the dispersion, the nonlinear part of the wavelength solution, or in the PSFlet shapes over several calibration sequences and cooling cycles.  Wavelength solutions based on more recent calibration sequences have not improved the extracted data cubes so long as a contemporaneous narrowband flat is available (see the following section for details).  We will add updated calibration files to the pipeline if and when they prove necessary.

\subsection{Narrowband Flats}

We use the long calibration sequences described above to measure the lenslet PSFs as a function of both wavelength and position, and to compute the wavelength solution.  This wavelength solution changes as CHARIS thermally cycles and as it is craned onto and off of its bench.  Small shifts due to motion of the optics or lenslet array change the position and orientation of the wavelength solution, but do not have a measurable impact on its nonlinear component.  We use a single narrowband flat for each night to compute these shifts and adjust the wavelength solution accordingly, assuming its nonlinear component to remain fixed.

Our strategy for these narrowband flatfield images is to uniformly illuminate CHARIS's lenslet array using the halogen flat-field lamp in front of AO188, the system optically upstream of SCExAO.  One of three narrowband filters within SCExAO creates a flatfield image that CHARIS cannot resolve spectroscopically.  These filters, at 10~nm width, are somewhat broader than our tunable source ($\sim$5~nm); the 1200~nm $J$-band light ($R \sim 120$) is marginally resolved by CHARIS in its $J$-band mode ($R \sim 80$).  All filters are completely unresolved in CHARIS's broadband mode.  

We process a narrowband flat using the same procedure as for the full calibration sequence above, except that we use the existing wavelength solution as a starting guess for the spot locations.  We perform a grid search to find the approximate offset and then iterate using Powell's method within the minimization function of \verb|scipy.optimize|.  We keep the offset and linear terms of this new solution, and replace the higher-order terms with their values in the main wavelength solution.  These refinements to the wavelength solution are all that we extract from the narrowband flats; they are the only aspects of CHARIS calibration data that are unstable from run to run.  Because CHARIS sits on Subaru's Nasmyth platform and because we designed the imaging relay to be thermally stable (with aluminum mirrors and an aluminum bench), the location of the PSFlets is very stable within a night.  Project 1640 and GPI are both mounted at the Cassegrain focuses of their respective telescopes; their PSFlets shift due to flexure as the telescope elevation angle changes \citep{Zimmerman+Brenner+Oppenheimer+etal_2011,Wolff+Perrin+Maire+etal_2014}.

We include the script \verb|buildcal| with the CHARIS software for the purpose of building all of the necessary calibration files.  The user runs \verb|buildcal| with the raw file for the narrowband flat as a command-line argument.  The software will then compute the calibrations and write all of the files to the directory from which \verb|buildcal| was run.  Some of these files are specific to optimal extraction and $\chi^2$ extraction, the two main spectral extraction techniques used by the pipeline.  These files are described together with the methods themselves in Sections \ref{subsec:OptimalExtraction} and \ref{subsec:chisqextract}.  A summary of the wavelength calibration steps and required products for each extraction technique is shown in Figure~\ref{fig:Wavecal}.

\begin{figure*}
	\centering
        \includegraphics[width=\linewidth]{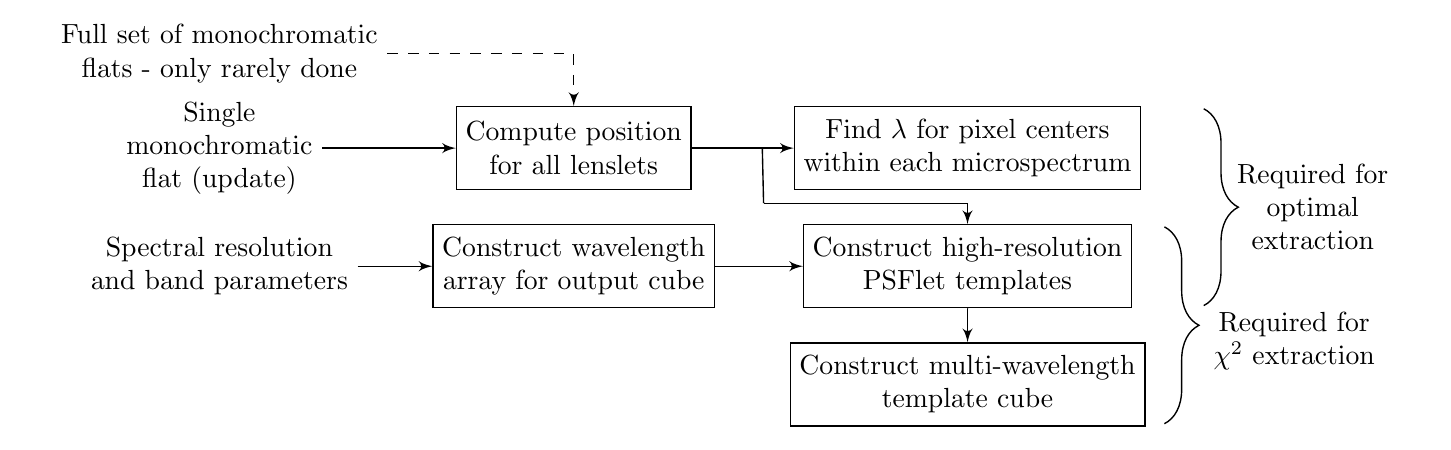}
	\caption{Wavelength calibration steps and required products for the two main reduction methods. A single monochromatic flat for each observing mode is the only data required by the CHARIS user to update all of these products and proceed to cube extraction. These steps are implemented in the script \texttt{buildcal}.}
	\label{fig:Wavecal}
\end{figure*}

\subsection{Masking and Flatfielding}

Flatfielding an ordinary image divides by a single sensitivity for each pixel; that sensitivity is a combination of illumination by the optics and the quantum efficiency of the pixel. In an IFS like CHARIS, these two components of the flatfield separate into a detector flat, measuring the pixel-to-pixel sensitivity differences, and a lenslet flat, measuring differences in the illumination and transmission of the various lenslets.  

We construct the pixel flat from early ramps taken before CHARIS was effectively baffled, when the lenslet array was (relatively) uniformly illuminated.  This flatfield does include artifacts from the nonuniform illumination.  We therefore apply a high-pass filter, preserving lenslet-to-lenslet variations, but removing the slow variations in illumination across the chip.  Individual CHARIS microspectra are $\sim$3$0 \times 6$ pixels in size; our high-pass filter is a two-dimensional Gaussian with $\sigma=10$\,pixels (full width at half maximum 23.5 pixels) and removes power on scales significantly larger than this.  

We use pre-baffling images, together with later images where the background count rate is very low, to identify bad pixels.  These are either hot, with a very high dark current, or they respond to light much more weakly than their neighbors.  Later images effectively identify the hot and warm pixels, while our method of constructing the pixel flat identifies pixels that are not light sensitive.  We flag a pixel as ``bad'' if it has a dark current 15 sigma above the read noise of its neighbors in a series of long exposure dark frames (i.e.~above $\sim$0.4\,$e^-$\,s$^{-1}$), or if its sensitivity in our high-pass-filtered pixel flat is below 80\%.  We flag 0.6\% of pixels as bad in this way; they are clustered in groups large and small across the detector.  

Our second step is to construct a lenslet flat.  For this step, we return to our full calibration sequences from which we derived a wavelength solution and measured the high-resolution lenslet PSFs.  Having extracted these, we reconstruct the monochromatic spot pattern expected for a uniformly sensitive detector and a uniformly illuminated lenslet array.  We then scale this ideal spot pattern to the pixel flat described above and determine the best fit lenslet-by-lenslet amplitude of the PSFlets.  Because the spot pattern is monochromatic, the spots are very well separated and crosstalk between neighboring lenslets may be safely ignored (see Figure \ref{fig:calsequence}).  

We thus obtain a lenslet flat for each wavelength in our sequence.  The flats are consistent with one another within the observing mode, as is expected for illumination of the field by nearly all reflective optics.  This also indicates that the filters, which lie immediately behind the lenslet array, have transmission curves that are nearly spatially uniform.  We median-combine the flats to create our lenslet flat for each observing mode.  We have also verified that these lenslet flats are consistent with those that we would derive from our narrowband flatfield images, described in the following subsection.

\begin{figure*}
    \includegraphics[width=\linewidth]{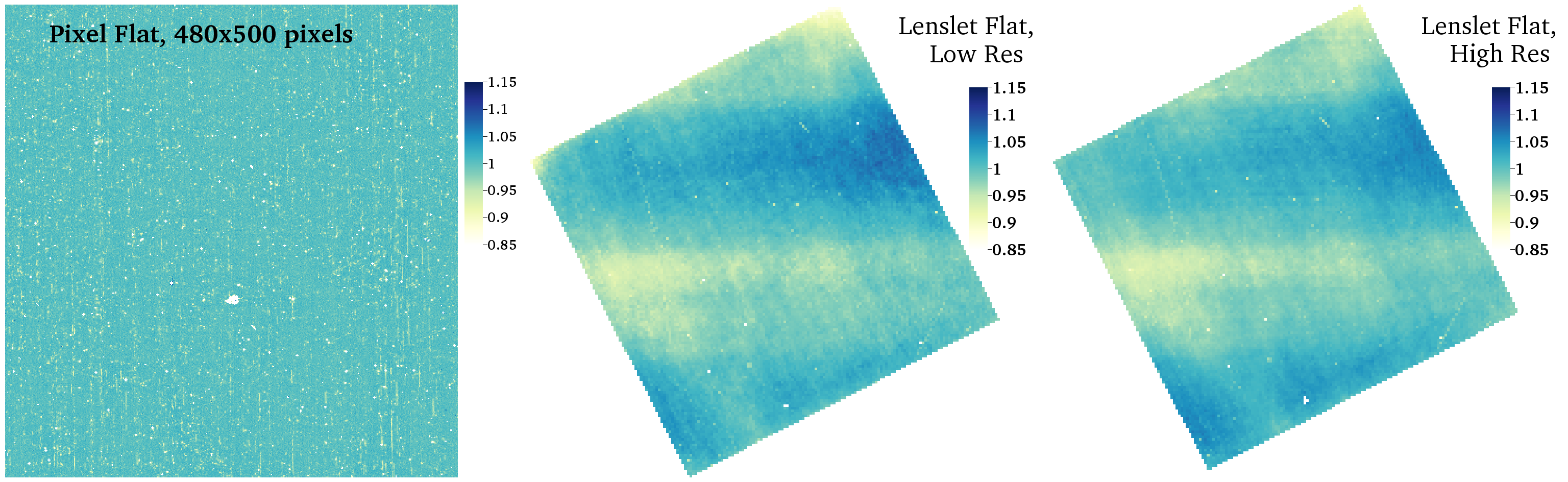}
    \caption{Left panel: a subregion of the pixel-based flatfield constructed before effective baffling was installed.  We used a high-pass filter to remove the illumination pattern.  Middle and right panels: lenslet flatfield images constructed by comparing the monochromatic spot pattern expected for perfectly uniform throughput, and the spot patterns actually observed.  The flatfields shown are averaged over wavelength and are very nearly consistent between the low-resolution and high-resolution modes.}
    \label{fig:flatfields}
\end{figure*}

Figure \ref{fig:flatfields} shows the lenslet flat and a subregion of the pixel flat.  The pixel flat is the same for all observing modes (it was measured using white light while the detector was poorly baffled), while there is, in principle, a different lenslet flat for each observing mode.  In practice, the lenslet flat is nearly the same in all modes due the almost exclusive use of reflective optics throughout CHARIS and the AO systems.  The pixel sensitivity generally varies by just a few percent from one pixel to the next.  The lenslet flat, apart from a few poorly illuminated (or effectively opaque) lenslets, has typical variations of $\sim$20\% across the array.

\section{Cube Extraction}
\label{sec:CubeExtraction}

Once we have constructed all of the calibration materials, extracting a data cube is relatively straightforward.  CHARIS includes two main techniques to perform the extraction: optimal extraction \citep{Horne_1986}, and $\chi^2$ extraction.  We describe our implementation of each below.  The CHARIS pipeline also implements a simple aperture photometry-based extraction.  However, this algorithm is unable to account for bad or noisy pixels and has no advantage, either in performance or run time, over optimal extraction.  

\subsection{Optimal Extraction}
\label{subsec:OptimalExtraction}

Optimal extraction \citep{Horne_1986} computes the spectral intensity at each wavelength (i.e.~each row of pixels perpendicular to the dispersion direction) using both the measured shape of the line-spread function and the pixel-specific measurement errors to weight the pixels.  This extraction method has long been the standard approach for spectrographs \citep{Baranne+Queloz+Mayor+etal_1996,Cushing+Vacca+Rayner_2004,Bolton+Schlegel+Aubourg+etal_2012}.  We implement optimal extraction for CHARIS assuming Gaussian profiles with a wavelength- and lenslet-dependent width that we measure from our high-resolution PSFlets (Figure \ref{fig:hires_psflets}).  This is equivalent to computing the spectral intensity as the normalization of a one-dimensional Gaussian of known position and width and unit area.

We compute the position of each lenslet's microspectrum as part of the calibration process, but the wavelengths that correspond to integer pixels along the spectrum differ.  Each microspectrum is $\sim$30 pixels long in the dispersion direction.  We obtain $\sim$30 ($\lambda,\sigma,x,y$) quadruples for each lenslet, where the positions are integers in the dispersion direction $y$ and floating point numbers in the perpendicular direction $x$ (which runs along the center of the microspectrum).  The wavelength-dependent width $\sigma$ of the microspectrum is given by computing the second moment of the flux along a line passing through the center of the corresponding PSFlet (Figure \ref{fig:hires_psflets}).  Because the PSFlets are not circularly symmetric, the actual profile will include variable contributions from the diffraction spikes at other wavelengths and will differ depending on a lenslet's spectrum.  Optimal extraction takes the ($\lambda,\sigma,x,y$) quadruples for each lenslet (which are computed and saved as part of the calibration step) and uses a weighted sum to calculate a corresponding spectral intensity.

Optimal extraction returns the spectral intensity at the wavelengths corresponding to a given microspectrum's sampling on the detector.  Each microspectrum has its own exact spectral resolution and subpixel sampling, and is therefore defined on its own native wavelength array.  It is possible to extract the data without interpolating onto a common wavelength array, but it is difficult to visualize and manipulate such data.  CHARIS' software therefore returns a cube in which these microspectra are interpolated onto a common wavelength array at the spectral resolution requested by the user.  The resulting data cube couples neighboring wavelengths because of this interpolation and also because the extracted microspectra are convolved with the line-spread function (a PSFlet's extent along the dispersion direction of a microspectrum).  

The output of optimal extraction is a pair of data cubes: one for the spectral intensities, and one for their errors.  The code could easily be modified to return cubes at CHARIS' native wavelength sampling (which differs for each lenslet).  In that case the software would return three cubes: one for the spectral intensities, one for their errors, and one for the native wavelengths.  

The CHARIS software also has the ability to perform a na\"ive aperture extraction using unit weights perpendicular to the dispersion direction.  This approach is currently used by SPHERE and GPI \citep{Perrin+Ingraham+Follette+etal_2016,Pavlov+Moller-Nilsson+Feldt+etal_2008,Mesa+Gratton+Zurlo+etal_2015}.  For CHARIS, aperture extraction produces cubes with slightly more noise (depending on the aperture) and lacks an ability to treat errors and bad pixels: bad or noisy pixels must be replaced with guesses from their neighbors.  We do not attempt to replace bad pixels in this case but simply set them to zero.  A more careful implementation of aperture photometry would produce a result no better than that from optimal extraction, and would save a negligible amount of computational cost.

\subsection{$\chi^2$ Extraction} \label{subsec:chisqextract}

CHARIS' software implements a $\chi^2$-based extraction, fitting every microspectrum with a linear combination of narrowband spots.  Once the narrowband spot templates have been computed, it is relatively straightforward to extract the spectrum, i.e., the coefficients of their best-fit linear combination.  Computing the templates themselves, however, is more subtle.  The spectra on the detector are the spectra seen by the lenslets, and convolved with both the lenslet PSFs and the pixel response function.  An attempt to use monochromatic PSFlets to fit the microspectra will suffer from the fact that no real spectrum can be represented as the sum of delta functions.  Attempting to overcome this by extracting a spectrum at a resolution significantly higher than CHARIS' native resolution would cause severe problems with aliasing and covariance between neighboring wavelengths.  Our solution is to fit the microspectrum of each lenslet as a series of top-hat spectra in units of $I_\nu$, with the spectral resolution of this sampling chosen to be comparable to, or slightly higher than, CHARIS' intrinsic spectral resolution.

We construct a spot diagram for a narrow spectral range using the oversampled lenslet PSFs like those shown in Figure \ref{fig:hires_psflets}.  We use ten monochromatic spots to construct each narrowband spot, first scaling each monochromatic spot by the atmospheric and filter transmission appropriate to the very narrow range of wavelengths it covers.  In this way, our narrowband spectra are what we would expect for an astrophysical source with a perfectly flat $I_\nu$, average atmospheric transmission, the filter for the given observing mode, and achromatic optics in the rest of the system.  The correction is far from perfect, but does mitigate the effect of the atmosphere and filter on our recovered microspectra.

\begin{figure*}
    \centering
    \includegraphics[width=0.9\textwidth]{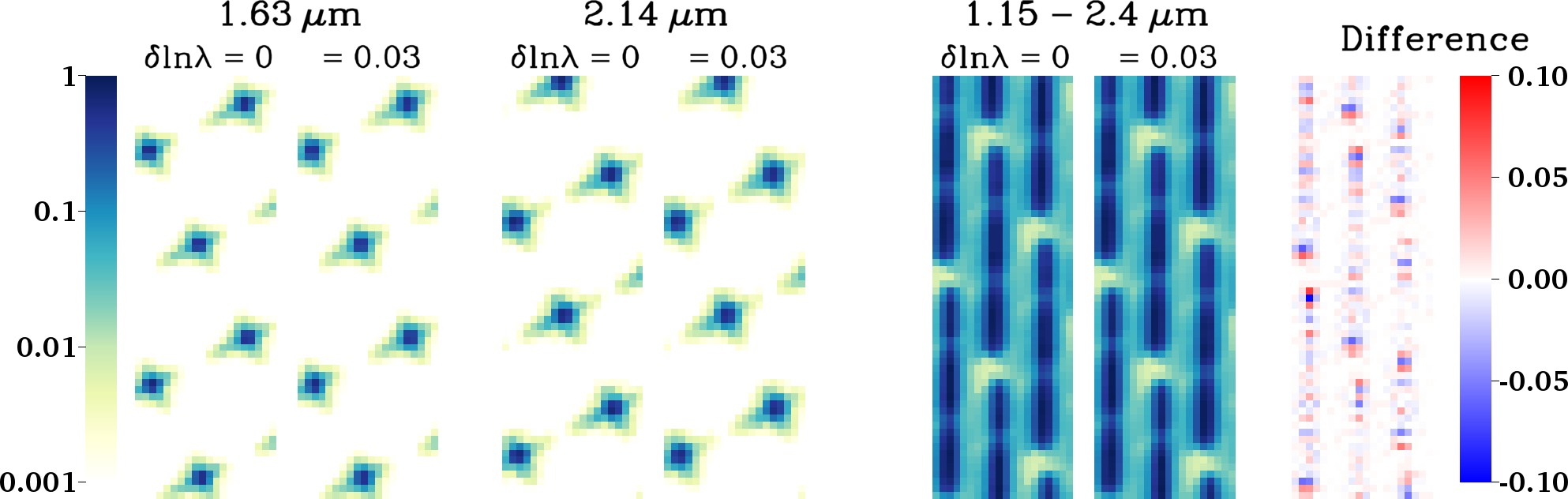}
    \caption{An illustration of the calibration products for $\chi^2$ extraction.  The panels labeled $\delta \ln \lambda = 0$ use monochromatic spots, while the panels labeled with $\delta \ln \lambda = 0.03$ account for the finite bandwidth separating the spectral measurements and also include wavelength-dependent atmospheric transmission.  The lenslet spots with $\delta \ln \lambda = 0.03$ are very slightly (almost imperceptibly) broader than those with $\delta \ln \lambda = 0$.  When modeling the microspectra of a source with constant $I_\nu$, the difference (right panel) between using monochromatic spots (third from right) and spots of finite bandwidth (second from right) is a high-frequency pattern with an amplitude around 10\% of the maximum intensity, peaking at the edges of the bandpass.}
    \label{fig:monochrome_polychrome}
\end{figure*}

Our calibration routine produces and saves a series of narrow, but not quite monochromatic, arrays of spots stepping through wavelength and built as described above.  Together, these spots cover the full wavelength array of a given observing mode with no gaps, and approximately correct for the chromatic throughput of the atmosphere and filter. Figure \ref{fig:monochrome_polychrome} shows these narrowband spot diagrams.  They are broader than those shown in Figure \ref{fig:calsequence}, but only slightly. 

Finally, we produce one last set of narrowband spots where we retain an oversampling by a factor of 5 in the direction perpendicular to the dispersion.  This allows us to use cross-correlation to find the appropriate offset for an individual CHARIS ramp even if a calibration data set is not available from that night (or if it proves a relatively poor match).  This takes a significant amount of disk space ($\sim$2~GB per calibration set), but enables the spectra to be located to $\sim$0.01~pixels and more accurately fit by our extraction routine.  We do not oversample in the dispersion direction because we currently have no way of getting a sufficiently accurate wavelength solution for an individual image without a narrowband calibration flat.  

Once all of the narrowband templates shown in Figure \ref{fig:monochrome_polychrome} have been computed and saved, $\chi^2$ spectral extraction is relatively straightforward.  We begin by cutting out a 7 pixel wide, $\sim$35 pixel long rectangle around each microspectrum.  We then fit each two-dimensional cutout with 20-25 narrowband spots, minimizing the squared residuals weighted by the pixels' inverse variance.  We use the singular value decomposition for this purpose, implementing it ourselves in Cython to enable lenslet-by-lenslet parallelization using OpenMP.  This approach naturally includes the errors on individual pixels and masks hot pixels, and it returns the spectral covariance matrix for each lenslet.  It also avoids any interpolation onto a common wavelength array as was necessary for optimal extraction.  A similar approach has been implemented for GPI \citep{Draper+Marois+Wolff+etal_2014,Ingram+Ruffi+Perrin+etal_2014}, but measurements of GPI's wavelength-dependent lenslet PSFs were insufficient for the algorithm to perform well.

Figure \ref{fig:chisq_resid} shows the performance of our $\chi^2$ extraction on the microspectra in two regions of a low-resolution image: one far from the star where read noise is important (upper panel), and a region in a bright speckle where it is negligible (lower panel).  The residuals from systematic errors in the PSFlet models are typically $\sim$5\% of the intensity.  As we discuss below in Section \ref{subsec:readnoise}, our models (center panels) include the correlated component of the read noise and the undispersed background.  This is visible in the top panel, where the left few columns of microspectra are in a noisy readout channel while the rightmost columns are in a much cleaner channel.

\begin{figure}
    \centering
\includegraphics[width=\basefigwidth]{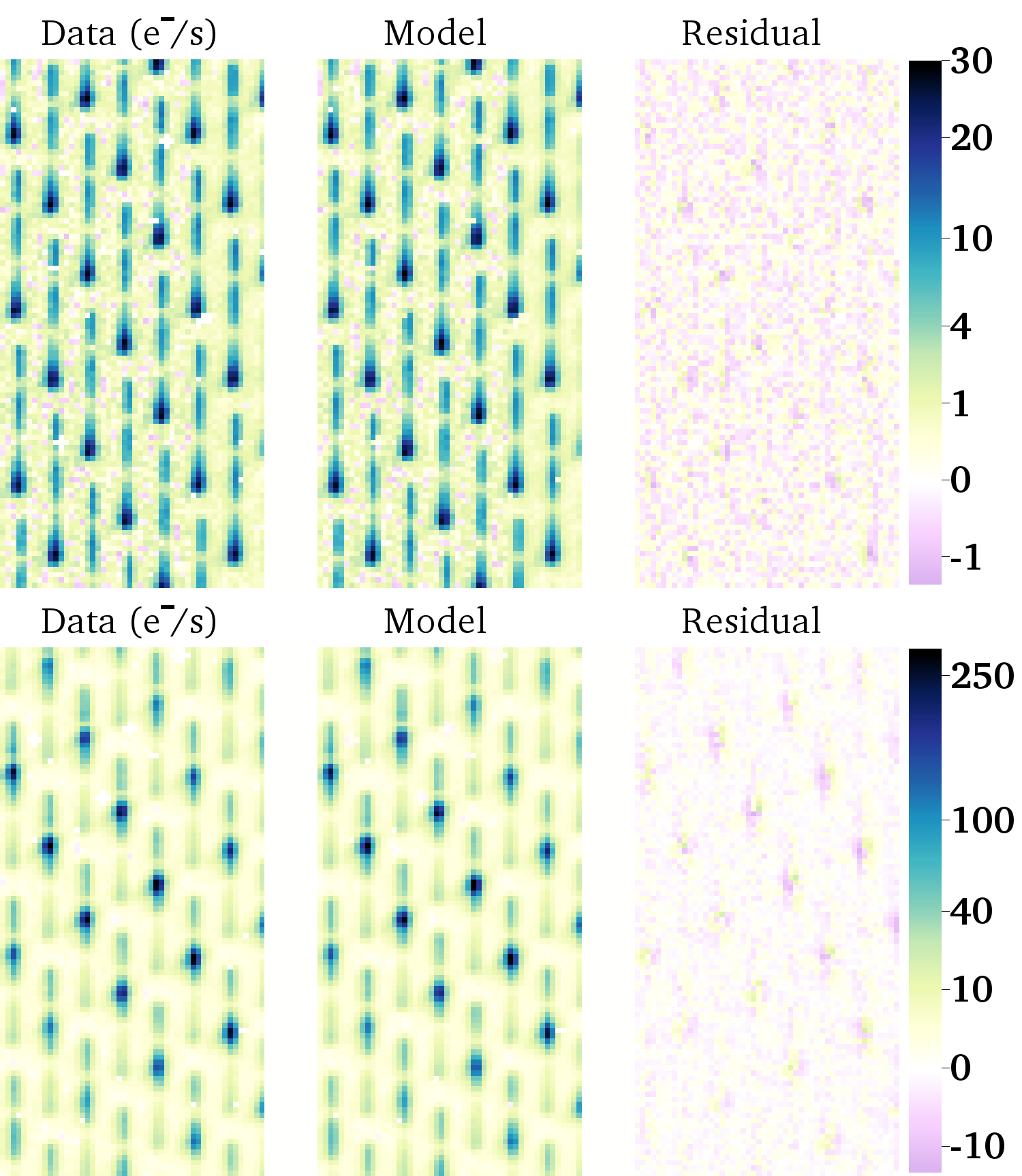}
\caption{The performance of our $\chi^2$ extraction algorithm on low-resolution microspectra in two regions of an image: one far from the star where read noise is significant (upper panel), and one in a bright speckle where it is negligible (lower panel).  We have used a square root stretch to more clearly show residuals.  Our model includes the undispersed background and the correlated component of the read noise, computed as described in Section \ref{subsec:readnoise}.  In the absence of read noise, the residuals are typically $\sim$5\% of the input data.}
\label{fig:chisq_resid}
\end{figure}

As an optional feature, the software can use the \mbox{PSFlet} template file that is oversampled in the direction perpendicular to the dispersion.  It uses cross-correlation over $32 \times 32$ subregions of the detector to fit a position-dependent subpixel shift in the locations of the spectra.  We adopt a range of prospective shifts spaced by 0.2 pixels, compute the cross-correlation in each case, and then fit a parabola to the three cross-correlation values nearest their minimum to obtain the exact subpixel offset.  We then use bilinear interpolation to compute an offset for each lenslet from the $32 \times 32$ cross-correlation offsets, and interpolate the oversampled PSFlet templates onto the appropriate pixel sampling.  This approach typically produces an offset of no more than a few tenths of a pixel, but noticeably improves the residuals in the two-dimensional microspectra.  The fit shown in Figure~\ref{fig:chisq_resid} includes this subpixel offset.

A $\chi^2$-based extraction has several features that differentiate it from optimal extraction and aperture photometry on the microspectra.  Because we fit the entire two-dimensional spectrum, it is trivial to also fit for and remove an undispersed background that is uniform over the microspectrum.  For microspectra that straddle two readout channels, we allow for the undispersed background to have different values in the two channels.  This fitting of an undispersed background is an optional setting in the software's configuration.

\begin{figure*}
    \includegraphics[width=\linewidth]{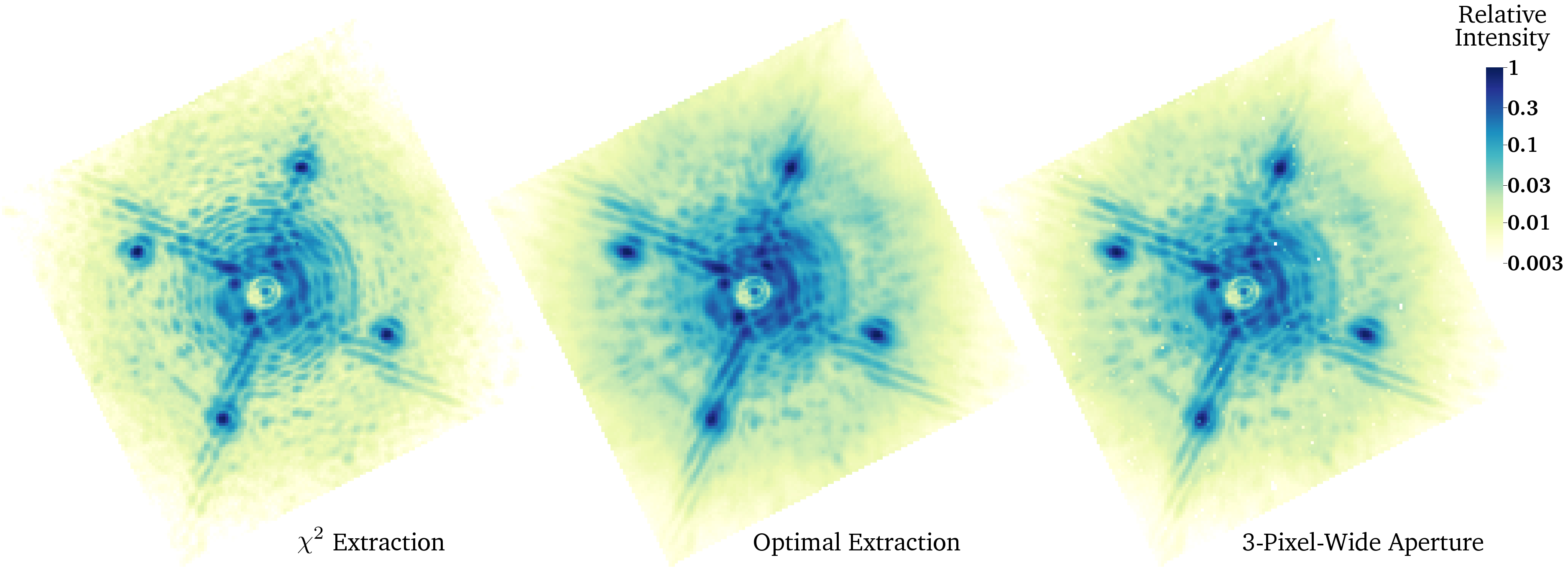}
    \caption{A single $H$-band slice through the same cube extracted three ways: $\chi^2$ extraction (left panel, Section \ref{subsec:chisqextract}), optimal extraction (middle panel, Section \ref{subsec:OptimalExtraction}), and an extraction using an unweighted 3-pixel-wide aperture (right panel).  A $\chi^2$ extraction intrinsically deconvolves the microspectra and the line-spread function, resulting in a more monochromatic image.  Speckles in the optimal extraction slice are radially extended because of both the convolution with the line-spread function and interpolation onto a common wavelength array.  The extraction using an unweighted aperture (right panel) is similar to optimal extraction but with slightly more noise and an inability to handle bad pixels.}
    \label{fig:lstsq_optext}
\end{figure*}

An important feature of the $\chi^2$ extraction is that it automatically attempts to extract the intrinsic source spectrum: it performs a deconvolution with the instrumental line-spread function.  This deconvolution results in a negative covariance between neighboring spectral bins and a superficially noisier cube than that produced by optimal extraction.  A small amount of smoothing in the spectral dimension, i.e.~a reapplication of the convolution with the line-spread function, removes this extra apparent noise.  The effect of the line-spread function is apparent in a single slice through a data cube, shown in Figure \ref{fig:lstsq_optext}.  With optimal extraction, the speckles are radially extended due to contributions from a range of wavelengths.  This arises both from the line-spread function and from the fact that we interpolate all of our microspectra (each of which has a different native wavelength sampling) onto a common wavelength array.  The speckles in a $\chi^2$-extracted cube show a negligible radial extent beyond that of the monochromatic PSF incident on the lenslet array.

A $\chi^2$ extraction naturally enables the removal of CHARIS' spectral crosstalk.  In the calibration step, we measured the lenslet PSFs (shown in Figure \ref{fig:hires_psflets}) out to a radius of $\sim$7 pixels, slightly larger than the $\sim$6 pixels horizontally separating the centers of the microspectra.  While we only fit the microspectra over a rectangle 7 pixels wide, our model microspectrum extends out to a box roughly twice as wide.  Subtracting all of the microspectra results in a small, negative residual as the broader boxes remove a few photons from the neighboring spectra.  We iterate one time on the cube to remove crosstalk.  In practice, the effects of crosstalk in CHARIS are extremely minor due to the spacing of our microspectra and our use of pinholes on the back of the lenslet array.  The correction from an iteration to remove crosstalk is generally $\ll$1\%.

\subsection{Residual Intensity and Read Noise Suppression} \label{subsec:readnoise}

A particularly important feature of our $\chi^2$ extraction is that it produces a two-dimensional model of the entire detector readout.  Subtracting this model from the actual ramp produces a residual image that, in the limit of perfect models of the microspectra, is pure noise.  Residuals from actual CHARIS ramps do show some systematics in the centers of the microspectra but are noise-dominated over much of the detector, both between microspectra and in the less-illuminated lenslets.  CHARIS' H2RG detector has an excessive amount of read noise correlated between the readout channels, as discussed in Section \ref{subsec:readnoiseprop} and shown in Figure \ref{fig:readnoisesamples}.  We use our measurement of a residual intensity to model and remove the correlated component of the read noise, achieving a suppression approaching that shown in the lower-right panel of Figure \ref{fig:readnoisesamples}.  

We fit for the correlated read noise as two patterns, one shared by the even readout channels and a second shared by the odd channels.  We also fit for a scalar coupling between each channel and the appropriate noise pattern. We compute the correlated read noise using a trimmed mean over a user-specified fraction, by default 70\%, of the pixels with the smallest fitted intensity relative to the read noise, i.e., the pixels where the residual is most dominated by the read noise.  Each element of the correlated read noise pattern has sixteen realizations on the detector (half of the 32 readout channels); this approach generally gives at least $\sim$10 pixels from which to calculate the pattern.  We then scale the correlated read noise by our fitted couplings and subtract it from each readout channel.

We find that our approach provides excellent suppression of the correlated component of the read noise while avoiding the addition of systematics back into the data.  Figure \ref{fig:readnoisedemo} shows an example of a slice through an extracted cube with and without suppressing the read noise and fitting out an undispersed background.  The image was taken through a neutral density filter to prevent saturation of the central star, resulting in low signal-to-noise ratios over much of the image.  The correlated read noise was severe on the date shown, but it is mostly removed by our algorithm.  

\begin{figure}
    \centering
    \includegraphics[width=\basefigwidth]{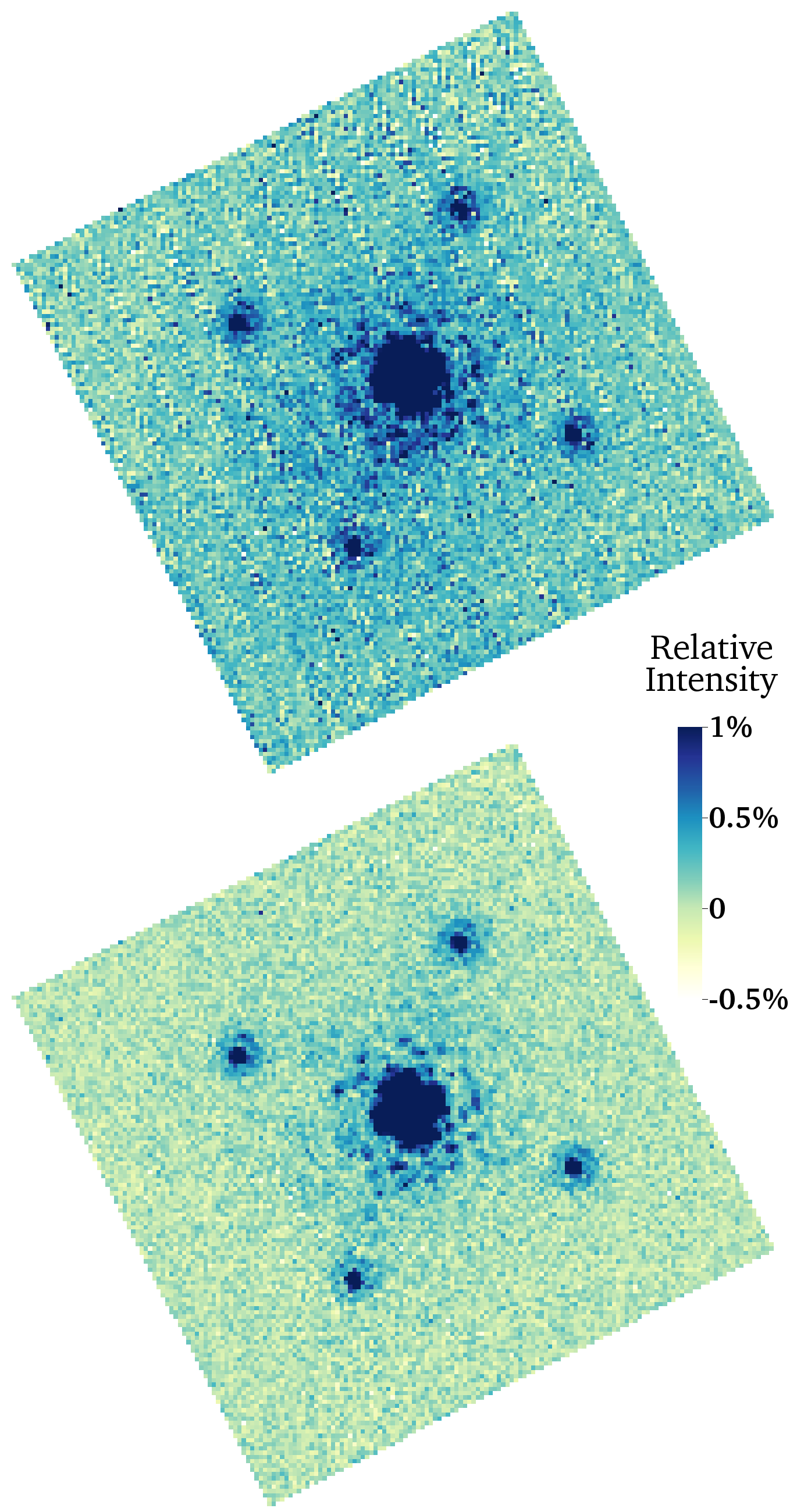}
    \caption{A single $H$-band slice through a low-intensity cube extracted with our $\chi^2$-based algorithm without (top panel) and with (bottom panel) fitting out the correlated read noise and an undispersed background.  The scales are the same on both images.  Low-background regions are about 55\% as noisy in the lower image as in the upper image.  The noisier readout channels are visible as stripes in the slice without read noise suppression.  }
    \label{fig:readnoisedemo}
\end{figure}

We also enable the suppression of correlated read noise when using optimal extraction or aperture photometry.  In this case, we use $\chi^2$ extraction but save the read noise and (optionally) the undispersed background rather than the data cube.  We then subtract the fitted read noise before performing the requested extraction algorithm.  This approach has the same effectiveness at removing correlated read noise as simply using $\chi^2$ extraction.

Our approach to read noise suppression works because of CHARIS' redundancy.  While the detector has $2048\times 2048$ pixels, we extract only $\sim$22 spectral measurements for each of our $\sim 135 \times 135$ illuminated lenslets, using $\sim$10 pixels on average to fit each spectral measurement.  The problem is sufficiently overconstrained that we can also fit two $2048 \times 64$ noise patterns without facing significant degeneracies.

\subsection{Background Subtraction} \label{subsec:bgsub}

CHARIS ramps have a background overwhelmingly composed of light leaks and thermal photons in the $K$-band--the true dark current is negligible.  The software does have the ability to subtract a background count rate from a two-dimensional ramp (i.e.~a matched dark).  This is common practice for near-infrared IFSs including OSIRIS \citep{Larkin+Barczys+Krabbe+etal_2006} and GPI \citep{Perrin+Ingraham+Follette+etal_2016}.  However, it is only worthwhile for CHARIS if a very high signal-to-noise ratio dark frame is available in the same instrument configuration as the science data.  Shot noise and read noise from the background frame will add to each image in an observing sequence, and because the background has a single realization of the noise, it will add coherently to all images.  The signal-to-noise ratio in the background must be substantially higher than in each science frame (i.e.~the integration time must be longer) to avoid this becoming a problem.  Matching exposure times will result in the background subtraction contributing $\sim$half the noise to each science frame, and most of the noise to a stack of science frames (because the same realization of noise is added to each frame).

The complexity of subtracting a background image is largely due to the fact that the background contains both a dispersed and an undispersed thermal component.  The dispersed component is overwhelmingly due to longer wavelengths, and is orders of magnitude lower with a filter in place that blocks $K$-band light.  Microspectra from the thermal background are clearly visible on the detector in the low-resolution and $K$-band modes, but are nearly invisible in the $J$ and $H$ bands or with the neutral density filter in place.  Figure \ref{fig:thermal_bkgnds} shows these backgrounds over part of the detector in the low-resolution, $K$, and $H$ bands.  The dispersed component dominates the background in the low-resolution and $K$-bands, while the $H$-band shows a relatively uniform background of $\sim$0.2~$e^-\,{\rm s}^{-1}$.  The microspectra in $K$-band are dispersed about four times as much as the low-resolution spectra, resulting in $\sim$$\frac{1}{4}$ of the peak intensity.

\begin{figure}
    \centering
\includegraphics[width=\basefigwidth]{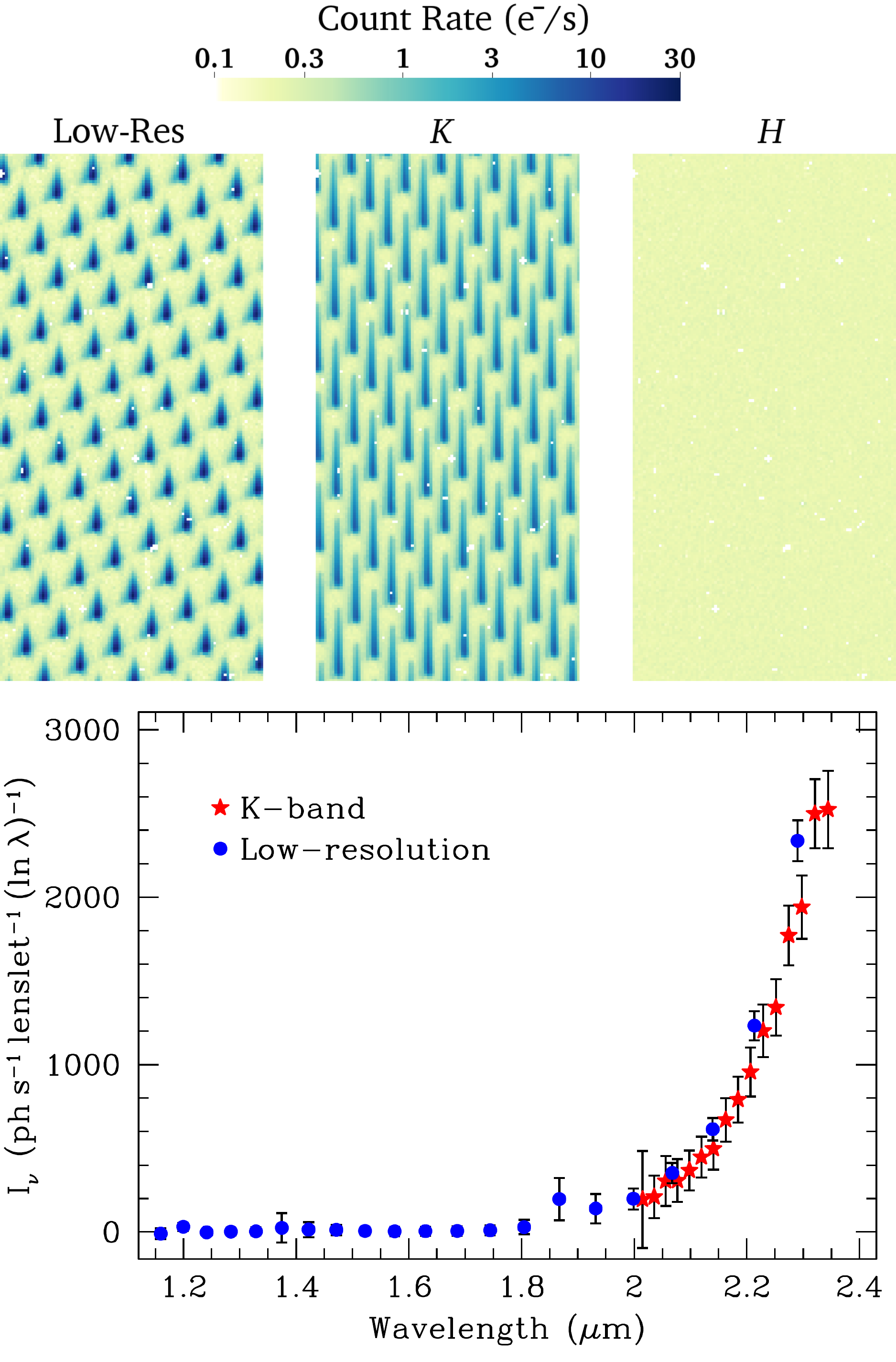}
\caption{Top panel: thermal backgrounds in the low-resolution, $K$, and $H$-band modes.  Bottom panel: spectral intensity of the dispersed background in low-resolution and $K$-band modes showing its mean and standard deviation across the lenslet array.  The background is composed of both a thermal component incident on the lenslet array and dispersed into microspectra by the prism, and a nearly uniform, undispersed component.  The dispersed component, which dominates the low-resolution and $K$-band backgrounds, is mostly from $K$-band light and is blocked by the $J$- and $H$-band filters.  }
\label{fig:thermal_bkgnds}
\end{figure}

Any change in the alignment of the lenslet array will change the locations of the thermal microspectra on the array.  The CHARIS extraction script can handle shifts in the location of the microspectra by applying a subpixel offset to the template PSFlets, but it cannot use this to shift the two-dimensional array of thermal microspectra.  Any mismatch, even if it is just a fraction of a pixel, will degrade the ability of $\chi^2$ extraction to fit the background-subtracted microspectra.  Unfortunately, we lack the hardware (e.g.~a mechanism to block the lenslet pinholes) to measure only the undispersed background.  

It is easy to degrade data by subtracting a poor two-dimensional background.  This is especially true of low-resolution data taken with the neutral density filter.  Exposures with this filter are usually longer integrations to achieve good signal-to-noise ratios, and therefore require long background images to perform a two-dimensional subtraction without adding noise.  Subtracting a background taken with the broadband (rather than neutral density) filter results in a dispersed thermal background orders of magnitude too high and a negative inferred sky background in the $K$-band.

In the absence of a matched, high signal-to-noise ratio, two-dimensional background, the software does have the ability to subtract an undispersed background lenslet-by-lenslet as part of a $\chi^2$ extraction (Section \ref{subsec:chisqextract}), or combined with read noise suppression (Section \ref{subsec:readnoise}).  The dispersed background will still be included in the final data cube (though it can easily be removed by standard post-processing algorithms).  

\subsection{The Extracted Cube}

The script \verb|extractcube| produces a FITS file with the data cube (the intensity at each lenslet and at each wavelength) as HDU~1 and the inverse variance of the data cube (the diagonal of the covariance matrix if using $\chi^2$ extraction) as HDU~2.  The first HDU, HDU~0, consists only of a header with key information about the observation and the data reduction.  The last HDU, HDU~3, contains the original header of the raw reads.  

The cube is defined on a logarithmic wavelength array.  It is given by the nonstandard header keywords \verb|LAM_MIN|, \verb|LAM_MAX|, and \verb|DLOGLAM|, with 
\begin{equation}
    \lambda_i = {\tt LAM\_MIN} \times \exp \left[ i \cdot {\tt DLOGLAM} \right]
\end{equation}
for $i = 0, \ldots, {\tt NLAM}$; the spectral resolution of the extracted cube is $R = 1/{\tt DLOGLAM}$.  The wavelength scale is also given by the standard FITS keywords  \citep{Greisen+Calabretta+Valdes+etal_2006}, with \verb|CTYPE3|$\,=\,$\verb|'AWAV-LOG'|.  We use a logarithmic scale because of CHARIS' nearly wavelength-independent dispersion: each pixel in the dispersion direction is a nearly constant increment in $\log \lambda$.  A uniform spacing in wavelength would result in aliasing problems at the short wavelength end of the spectrum.

With $\chi^2$ extraction, the data cube is never interpolated except over bad spectral measurements (which in any case have their inverse variance set to zero).  With aperture photometry or optimal extraction, each lenslet's microspectrum has been interpolated to place it on a common wavelength array.  In any of these cases the inverse variance does not capture the full spectral covariance.  Still, it does capture lenslet-to-lenslet variations in the quality of the cube, and prevents bad measurements from corrupting the images in post-processing.

\section{Software Parameters and Settings} \label{sec:settings}

In order to reduce CHARIS data, we provide a software package that accomplishes the two required tasks: wavelength calibration (through the script \verb|buildcal|, see Section~\ref{sec:calib}), and cube reduction (the script \verb|extractcube|, see Section~\ref{sec:CubeExtraction}). The wavelength calibration only needs to point to a monochromatic flat taken as close as possible to the actual date the science data were taken, ideally the same night. The user can choose whether or not to compute over-sampled PSFlets, which are required to fit for sub-pixel shifts in the ramp as described in Section~\ref{subsec:chisqextract}.

The cube extraction script, \verb|extractcube|, has several parameters that the user can set; these will change the photometry and morphology of the reduced spatio-spectral datacube as described in Section~\ref{sec:CubeExtraction} and Figure~\ref{fig:lstsq_optext}. The set of parameters is presented in Table~\ref{tab:parameters} and can be edited within a \verb|.ini| text file that is read by the extraction software.

\ifspie
\else
\begin{deluxetable*}{lccl}
\tablewidth{0pt}
\tablecaption{Extraction Software Parameters}
\tablehead{	
    \colhead{Parameter} &
    \colhead{Default} &
    \colhead{Allowed Values} &
    \colhead{Description} \\
    }
\startdata
\multicolumn{4}{c}{\textit{Ramp Parameters}}		\\
\hline
${\tt read\_0}$	&	1	& $[1,\, N_{\rm reads} - 1]$ &	First read in the ramp to use. \\
${\tt read\_f}$	&	${\tt None}$	& $ [{\tt read\_0} + 1, \, N_{\rm reads}]$ or ${\tt None}$ &	Last read to use.  If ${\tt None}$, ${\tt read\_f}$ will be the final read.	\\
${\tt gain}$	&	2	&	$[0, \, \infty)$ & Detector gain, $e^-$/count, used to compute photon noise.	\\
${\tt noisefac}$	&	0	& $[0,\, 1)$ & Additional fractional error in count rate to account for systematic errors in \\
& & & $\qquad$ PSFlet models.	Suggested values: 0.05 for $\chi^2$ extraction, 0 otherwise.  \\
${\tt saveramp}$	&	${\tt False}$	&	${\tt True}$, ${\tt False}$ & Save the ramp as its own image?	\\
\hline					
\multicolumn{4}{c}{\textit{Calibration Parameters}}		\\
\hline
${\tt calibdir}$	&	\dots	& string file path &    Directory containing the files created by ${\tt buildcal}$	\\
${\tt bgsub}$	&	${\tt False}$	&	${\tt True}$, ${\tt False}$ & Subtract a background/thermal ramp?	If {\tt True}, {\tt buildcal} must have been \\
 & & & $\qquad$ called with background ramp(s).  Should be {\tt False} for ND frames. \\
${\tt mask}$	&	${\tt True}$	&	${\tt True}$, ${\tt False}$ & Mask bad pixels?	\\
${\tt flatfield}$	&	${\tt True}$	&	${\tt True}$, ${\tt False}$ & Apply the pixel and lenslet flatfields?  \\
${\tt fitshift}$	&	${\tt True}$	&	${\tt True}$, ${\tt False}$ & Fit for a subpixel shift of the microspectra?  \\
& & & $\qquad$ If ${\tt True}$, ${\tt buildcal}$ must have computed oversampled PSFlets.	\\
\hline					
\multicolumn{4}{c}{\textit{Extraction Parameters}}	\\
\hline
${\tt R}$	&	30 	& $[1, \, \infty)$	& Spectral resolution $R = \left(\delta \ln \lambda\right)^{-1} = \lambda/\delta \lambda$ of the extracted cube	\\
    &       &                   & Recommended: 30 for low res, 100 for high res \\
${\tt method}$	&	${\tt lstsq}$ & ${\tt lstsq}$, ${\tt optext}$, & Method used to extract cube: $\chi^2$ extraction, optimal extraction, \\
        &           & ${\tt apphot3}$, ${\tt apphot5}$	&	$\qquad$ or aperture photometry across 3 or 5 pixels.	\\
${\tt refine}$	&	${\tt True}$	&	${\tt True}$, ${\tt False}$	&	Iterate once to remove crosstalk?  Only used if method is ${\tt lstsq}$.	\\
${\tt suppressrn}$	&	${\tt True}$	&	${\tt True}$, ${\tt False}$	&	Use the method of Section \ref{subsec:readnoise} to remove read noise?	\\
${\tt minpct}$ & 70 & $(0,\,100)$ & Minimum percentage of pixels used to estimate read noise (only  \\
 &  &  & $\qquad$ used if ${\tt supressrn}$ is ${\tt True}$). \\
${\tt fitbkgnd}$ & ${\tt True}$	&	${\tt True}$, ${\tt False}$	& Fit and remove an undispersed background lenslet-by-lenslet? \\
${\tt smoothandmask}$	&	${\tt True}$	&	${\tt True}$, ${\tt False}$	&	Interpolate over bad lenslets to display a smooth cube? \\
& & & $\qquad$ The inverse variance of the bad measurements remains zero.	\\
${\tt saveresid}$	&	${\tt False}$	&	${\tt True}$, ${\tt False}$	&	Save the 2D residual from the ramp?  Only used if method is ${\tt lstsq}$.	\\
${\tt maxcpus}$	&	${\tt None}$	&	$[1-N_{\rm cpus},\, N_{\rm cpus}]$, or ${\tt None}$ & Maximum number of threads to use.  If ${\tt None}$, use all threads.  \\
& & & $\qquad$ If negative, it is the number of threads to reserve.  
\enddata		
\label{tab:parameters}
\end{deluxetable*} 

\fi

Many of the parameters in Table \ref{tab:parameters} are intended to be left as fixed.  Indeed, the only parameter that a typical user will need to modify from its default value is \verb|calibdir| (which must point to a local directory).  Some should never be changed from their default values, while others may be modified to suit a particular reduction or a user's particular needs.  Several of these parameters, however, allow the user to run a faster reduction, to change the treatment of the background, and to write intermediate data products. 

\ifspie
\else

\begin{deluxetable}{lcr}
\tablewidth{0pt}
\tablecaption{Suggested Parameter Settings}
\tablehead{	
    \colhead{Parameter} &
    \colhead{Fast \& Rough} &
    \colhead{Slower \& Better} \\
    \colhead{} &
    \colhead{Reduction} &
    \colhead{Reduction}
    }
\startdata
{\tt read\_0} & {\tt 1} & {\tt 1} \\
{\tt read\_f} & {\tt None} & {\tt None} \\
{\tt gain} & {\tt 2} & {\tt 2} \\
{\tt noisefac} & {\tt 0}  & {\tt 0.05} \\
\hline
{\tt bgsub}    & \tablenotemark{$\dagger$}{\tt False} & \tablenotemark{$\dagger$}{\tt False} \\
{\tt mask} & {\tt True} & {\tt True} \\
{\tt flatfield} & {\tt True} & {\tt True} \\
{\tt fitshift} & {\tt False} & {\tt True} \\
\hline
{\tt R} & \tablenotemark{$\dagger\dagger$}{\tt 30} or {\tt 100} & \tablenotemark{$\dagger\dagger$}{\tt 30} or {\tt 100} \\
{\tt method} & {\tt optext} & {\tt lstsq} \\
{\tt refine} & \ldots & {\tt True} \\
{\tt suppressrn} & {\tt False} & {\tt True} \\
{\tt minpct} & \ldots & {\tt 70} \\
{\tt fitbkgnd} & \tablenotemark{$\dagger$}{\tt False} & \tablenotemark{$\dagger$}{\tt True} \\
{\tt smoothandmask} & {\tt True} & {\tt True}
\enddata
\tablenotetext{$\dagger$}{See Sections \ref{subsec:bgsub} and \ref{sec:settings} for a discussion.  Never set both {\tt bgsub} and {\tt fitbkgnd} to {\tt True}.}
\tablenotetext{$\dagger\dagger$}{{\tt 30} in low-res mode, {\tt 100} in $J$, $H$, or $K$.}
\label{tab:demoparams}
\end{deluxetable}
\fi

Table \ref{tab:demoparams} lists two sets of parameters, one intended for a quick reduction and the other for a more detailed reduction that removes read noise and performs a $\chi^2$ extraction.  The parameters \verb|read_0|, \verb|read_f|, \verb|gain|, \verb|mask|, and \verb|flatfield| are the same in both reductions; we generally recommend that the user never change these values.  Changing either \verb|flatfield| or \verb|mask| to \verb|False|, for example, will save a negligible amount of computational effort.  In order to extract a cube with \verb|bgsub|, a thermal background must have been computed by \verb|buildcal|, while using \verb|fitshift| requires oversampling the PSFlets in \verb|buildcal|.  The parameter \verb|smoothandmask| is intended only for cosmetics: it replaces spectral measurements that are much noisier than their neighbors with values taken from a smoothed cube.  To avoid biasing the results, any modified intensities have their corresponding inverse variances set to zero.  Parameters marked by `\ldots ' are ignored, \verb|refine| because \verb|method| is not \verb|lstsq| and \verb|minpct| because \verb|suppressrn| is \verb|False|.  

One of the more problematic aspects of the reduction is background subtraction, discussed in Section \ref{subsec:bgsub}.  Subtracting a good match to the two-dimensional thermal background (comprising both the microspectra and undispersed background) can improve the reduction.  However, it is easy to degrade the final cube by subtracting a noisy background or one that is a poor match to either the thermal microspectra or the undispersed background.  For this reason, we generally recommend that a user set \verb|bgsub| to \verb|False|.  If the user has a background image matching the observing mode and with at least a factor of a few longer integration time than any of the science frames, then setting \verb|bgsub| to \verb|True| may improve the reduction.  If \verb|bgsub| is set to \verb|False|, then the undispersed background may be fit and removed by setting both \verb|fitbkgnd| and \verb|suppressrn| to \verb|True|, or by setting \verb|fitbkgnd| to \verb|True| with \verb|method| equal to \verb|lstsq|.  The user should never attempt to remove the background twice by setting both \verb|bgsub| and \verb|fitbkgnd| equal to \verb|True|.

\section{Software Performance and Scaling} \label{sec:performance}

The CHARIS software is written in Python and Cython \citep{Dalcin+Bradshaw+Smith+etal_2011}, and uses the extensive libraries available in the NumPy, SciPy \citep{VanDerWalt+Colbert+Varoquax_2011}, and Astropy \citep{Astropy_2013} packages.  In addition to the parallelization performed natively in these packages, we have used the multiprocessing module to parallelize the construction of model PSFlets and OpenMP to parallelize many of the Cython routines.  

We have tested the performance and scaling of the software on a compute server with two 12-core 2.5~GHz Intel Xeon processors.  We use \verb|maxproc| to set the maximum number of threads to allocate and test the two sets of parameters given in Table \ref{tab:demoparams}.  We use a 41-read ramp (about a 60-second exposure); a shorter exposure would require correspondingly less time to compute the ramp.  

\begin{figure}
    \centering
\includegraphics[width=\basefigwidth]{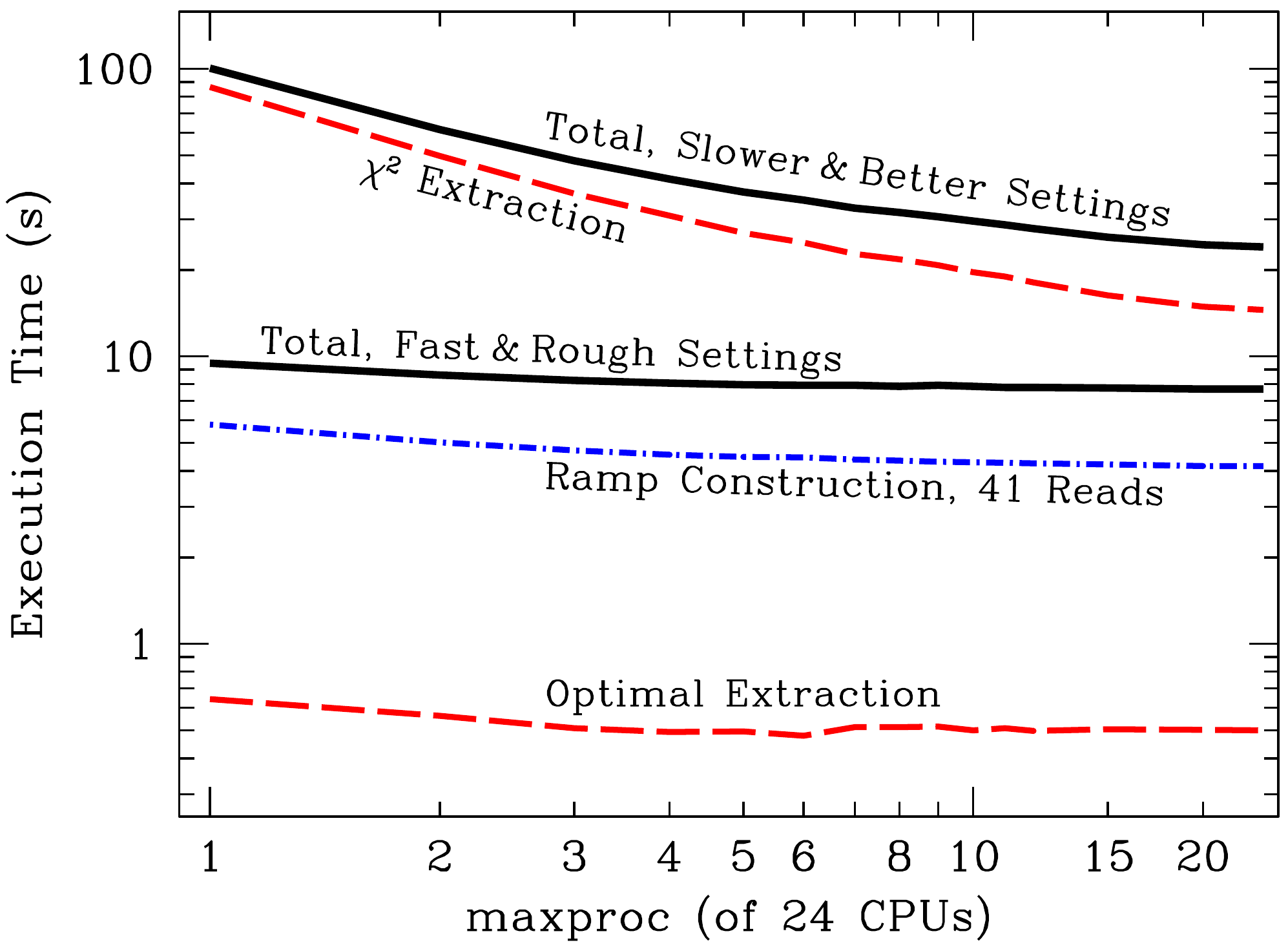}
\caption{Performance and scaling results on a server with 24 2.5~GHz Intel Xeon processors for the two suggested parameter sets in Table \ref{tab:demoparams}.  The $\chi^2$ extraction step scales reasonably well up to $\sim$10 CPUs and dominates the computational cost of our slower, better reduction.  The fast and rough reduction is dominated by the construction of the ramp (which requires a lot of I/O and scales poorly), overheads, and astropy computations of observational parameters.}
\label{fig:speed_stats}
\end{figure}

Figure \ref{fig:speed_stats} shows the results.  The software has a few seconds of overhead from the initial module imports and from astropy computations of important parameters of the observation (such as the parallactic angle).  The computation of the ramp requires a large amount of I/O and scales with the number of reads; using only the first ten of the 41 reads reduces the computational cost by about a factor of four.  For the fast and rough reduction, OpenMP provides only a slight performance gain and the cube extraction is a minor component of the total computational cost.  The slower and better reduction, however, sees large performance gains from parallelizing the $\chi^2$ extraction step, scaling reasonably well up to $\sim$10 CPUs.  This step requires the cube to be extracted three times: an initial extraction, a second extraction once a first estimate of the correlated noise has been removed (Section \ref{subsec:readnoise}), and finally after removing updated estimates of the correlated read noise and spectral crosstalk.  Setting \verb|refine| and \verb|suppressrn| to \verb|False| (which we do not recommend) would reduce the computational cost of the $\chi^2$ extraction step by nearly a factor of three.  Using optimal extraction but setting \verb|suppressrn| to \verb|True| would require at least two $\chi^2$ extractions (three if also computing an undispersed background).  In this case, the total computational time would be almost identical to the results shown here for the slower and better reduction with $\chi^2$ extraction.

\section{Discussion and Conclusions} \label{sec:conclusions}

In this paper, we have presented the data reduction pipeline for the CHARIS IFS.  The software begins with the raw detector reads, corrects for an artifact in the first read, and then reconstructs the count rate at each pixel using either up-the-ramp weighting or the full nonlinear response.  This approach handles saturation gracefully as long as there are at least a few reads before a pixel saturates, and it uses only those reads that are uncontaminated by bleeding from a neighboring saturated pixel.  Fitting the full nonlinear response to pixels that exceed $\sim$10\% of well capacity adds a negligible amount to the required runtime.  

There are two steps after constructing the ramp: first, all of the calibration files must be built from a narrowband flatfield image, and second, the data cube must be extracted.  Our calibration step uses a range of data products that we have built from early lab data and from detailed image sequences taken with a supercontinuum source and a tunable filter.  These products consist of the detector flatfield, the lenslet flatfield, a bad pixel mask, a reference wavelength solution (a wavelength-dependent mapping from lenslet to pixel coordinates), and reconstructed position- and wavelength-dependent lenslet PSFs.  We supply a script \verb|buildcal| with the software; \verb|buildcal| takes a narrowband filter from the same night as the data to be reduced and computes a perturbation to the wavelength solution and resampled PSFlets for each wavelength and position to be extracted.  

Once the calibration step has been completed, the script \verb|extractcube| extracts the data cube.  This script operates on a sequence of reads and requires a set of input parameters that the user may set using a \verb|.ini| file.  It first constructs the count rate at each pixel from the sequence of reads and computes basic parameters of the observation using keywords in the FITS header.  Next, \verb|buildcal| optionally computes a sub-pixel shift in the positions of the microspectra, and finally extracts the data cube.  

We implement three cube extraction algorithms: aperture photometry, optimal extraction, and $\chi^2$ extraction.  The GPI and SPHERE pipelines currently use aperture photometry as their primary extraction methods \citep{Perrin+Ingraham+Follette+etal_2016,Pavlov+Moller-Nilsson+Feldt+etal_2008,Mesa+Gratton+Zurlo+etal_2015}.  While a least-squares inversion technique has been implemented for GPI \citep{Draper+Marois+Wolff+etal_2014}, it has not yet equaled the performance of simple aperture photometry.  Project 1640's pipeline \citep{Zimmerman+Brenner+Oppenheimer+etal_2011} comes closest to our approach, but after reconstructing the lenslet PSFs, it uses a weighted sum with the PSFlet templates rather than a $\chi^2$ inversion.  This decision increases the coupling between neighboring wavelengths, as it effectively performs a second convolution with the line-spread function.

All of the aperture photometry and optimal extraction techniques require interpolation to place the microspectra onto a common wavelength array.  This interpolation couples neighboring wavelengths and degrades the data cube, while aperture photometry also requires bad data to be fixed.  Our $\chi^2$ extraction algorithm avoids all of these drawbacks, and also performs a deconvolution of the microspectra with the line-spread function, resulting in much more monochromatic slices through the data cube.  The $\chi^2$ algorithm produces a full two-dimensional residual map that allows us to estimate and remove correlated read noise, which results in a dramatic improvement to CHARIS data.  This algorithm also allows us to fit out and remove spectral crosstalk and an undispersed lenslet-by-lenslet background, and it produces a full covariance matrix for each lenslet's microspectrum (though we presently save only the diagonal elements).

The final data product of the CHARIS pipeline is an extracted $(x, y, \lambda)$ data cube with inverse variances for each spatial and spectral measurement.  This cube must subsequently be calibrated, using satellite spots induced by SCExAO \citep{Jovanovic+Guyon+Martinache+etal_2015} or some other method.  The pipeline currently saves an estimated transformation from lenslet to sky coordinates; we defer a full astrometric calibration to a future paper.  We caution against na\"ively interpolating our data cube, as doing so would mix spatial and spectral elements that could have very different uncertainties and will inevitably discard information in the cube.

The CHARIS software is written in Python and Cython, and is freely available on github at \url{https://github.com/PrincetonUniversity/charis-dep}.  It is parallelized, taking anywhere from a couple of seconds to run on a short ramp with a rough reduction to $\sim$100 seconds on a longer ramp with a detailed reduction using a single processor.  With a moderately powerful compute server, the most detailed reductions may be completed in $\sim$20--30 seconds per frame.  The CHARIS pipeline has a documentation page that we maintain at \url{http://princetonuniversity.github.io/charis-dep/}.  The data cubes that it produces are suitable inputs for high-contrast image processing, a topic that we will address in forthcoming work.

\acknowledgments{This work was performed in part under contract with the Jet Propulsion Laboratory (JPL) funded by NASA through the Sagan Fellowship Program executed by the NASA Exoplanet Science Institute.  This work was performed in part under a Grant-in-Aid for Scientific Research on Innovative Areas from MEXT of the Japanese government (Number 23103002).  This research is based on data collected at the Subaru telescope, which is operated by the National Astronomical Observatories of Japan.  The authors wish to recognize and acknowledge the very significant cultural role and reverence that the summit of Mauna Kea has always had within the indigenous Hawaiian community. We are most fortunate to have the opportunity to conduct observations from this mountain. }

\ifspie
\else
\bibliographystyle{apj_eprint}
\fi
\bibliography{mybib}

\ifspie
\listoffigures
\listoftables
\end{spacing}
\else
\fi

\end{document}